\documentclass[pre,twocolumn,groupedaddress,showpacs]{revtex4}
\usepackage{graphicx}
\usepackage{dcolumn}
\usepackage{amsmath}
\usepackage{bm}
\begin{document}
\title{Theory and simulation of the nematic anchoring coefficient}
\author{Denis Andrienko}
\author{Michael P. Allen}
\homepage{www.phy.bris.ac.uk/research/theory/simulation/}
\affiliation{H. H. Wills Physics Laboratory, University of Bristol, \\
Royal Fort, Tyndall Avenue, Bristol BS8 1TL, United Kingdom }
\date{\today}
\begin{abstract}
  Combining molecular simulation, Onsager theory and the elastic 
  description of nematic liquid crystals, we study the dependence of 
  the nematic liquid crystal elastic constants and the zenithal surface 
  anchoring coefficient on the value of the bulk order parameter. 
\end{abstract}
\maketitle
\section{Introduction}
\label{sec:intro}
The anchoring phenomenon is the tendency of a liquid crystal to orient 
in a particular direction when in contact with the container walls. 
The equilibrium director orientation set by the interaction of the 
liquid crystal with the aligning surface is called an 
\emph{easy orientation axis}.
The simplest, strong anchoring, assumption is that the director has a 
fixed orientation at the boundaries along the easy orientation axis.
However, it has been discovered that the  
coupling of the director with the orienting surface can be rather weak. 
This results in deviation of the surface director from the easy axis in response
to small perturbations.

On a phenomenological level, weak anchoring can be described by adding 
an appropriate \emph{surface potential} to the free energy of the system. 
Then minimization of the free energy functional gives both the equations 
for the director in the bulk and the appropriate boundary conditions
\cite{degennes.pg:1995.a}. 
The simplest form of the surface potential has been proposed by Rapini and 
Papoular \cite{rapini.a:1969.a}
\begin{equation}
f_{\rm{s}} = - \frac{1}{2} W(\bm{n} \cdot \bm{e})^2,
\label{rapini_potential}
\end{equation}
where the parameter, $W$, is termed an anchoring energy.  

Since then, numerous experimental methods have been used to 
measure the surface anchoring coefficient $W$ 
\cite{subacius.d:1995.a,yokoyama.h:1985.a,nastishin.y.a:1999.a};
its value has turned out to be extremely important for liquid crystal devices, 
i.e. displays, optical switches.
However, comparatively little work on systematic experimental
investigation of the anchoring phenomenon has been presented up till now.
From the available experimental data, one can say 
that the extrapolation length $\lambda=K_{33}/W$ is inversely 
proportional to the squared value of the bulk order parameter 
$\lambda \propto Q^{-2}$. 
Taking into account that the elastic constant $K_{33}$ is typically 
proportional to $Q^2$, this gives for the anchoring parameter $W\propto Q^4$ 
\cite{mertelj.a:2000.a}.

There have been several attempts to estimate the anchoring coefficient 
theoretically \cite{tjiptomargo.b:1988.a} and by combining molecular 
simulation with a local density functional approach 
\cite{stelzer.j:1997.b,teixeira.pic:2001.a}.
The main difficulty here is that one needs to know the direct pair correlation 
function of the nematic state, which is usually unknown, and must hence be
estimated with some uncontrolled approximations or assumptions (see, however,
Ref.~\cite{phuong.n.h:2001.a}). Moreover, 
it is often assumed that the director in the interface region is fixed
\cite{stelzer.j:1997.b}. However, it has already been noticed that 
these approximations may give incorrect results, by at least an order of 
magnitude, for example in the calculation of bulk  elastic constants 
\cite{phuong.n.h:2001.a}. 

Similar studies have also been done using lattice models.
The existence of subsurface deformations and effective intrinsic anchoring 
arising from the incomplete molecular interaction close to the surface 
has been studied using a hexagonal lattice approach \cite{skacej.g:1997.a}.
Monte-Carlo simulations of the Lebwohl-Lasher model have shown that
the extrapolation length is not in general equal to the ratio of 
the bulk to surface couplings \cite{priezjev.n:2000.a}. However, the results
were not sufficiently robust to determine the dependence of the 
anchoring energy on the order parameter.

The present work attempts to remedy the situation by combining the elastic
description with Onsager theory and Monte Carlo simulation results.
We study the dependence of the
elastic ($K_{33}$) and surface anchoring ($W$) coefficients on the 
liquid crystal state point, which is defined by the bulk value of the 
order parameter $Q$.

The paper is organized as follows. We define the geometry and derive director
profiles using elastic theory in section \ref{sec:elastic}. In section
\ref{sec:onsager} we discuss the Onsager approach which allows us to calculate
the single particle density of the bulk and confined systems, while section
\ref{sec:fluctuations} outlines the fluctuation approach to calculating elastic
coefficients. Section \ref{sec:simulation} gives details of the technique used
in Monte Carlo simulations and the results are summarized in section
\ref{sec:results}.

\section{Elastic description}
\label{sec:elastic}
The easiest way to obtain the surface anchoring coefficient $W$ is to create
a director deformation far from the surface. Then, measuring the response
of the director near the cell surface, or fitting the director profile
with a theoretically predicted profile, yields an estimate of the 
anchoring extrapolation length and ratios of the elastic constants.

Indeed, consider one of the possible geometries suitable for the measurement 
of the zenithal anchoring strength. Let the director have fixed orientation 
at the boundary $z = L$. The surface at $z = 0$ is assumed to provide 
homeotropic anchoring of strength $W$.
In the elastic description, deformations of the director field $\bm{n}$ 
are described by the total free energy \cite{degennes.pg:1995.a}
\begin{equation}
{\cal F}_{\text{el}} =
\int_V \, f_{\text{b}} \, dV + \int_S \, f_{\text{s}} \, dS \:.
\label{free_energy}
\end{equation}
Here $f_{\text{b}}$ is the Frank-Oseen elastic free energy density
\begin{eqnarray}
f_{\text{b}}&=&\frac{1}{2}\left\{ 
K_{11}\left( \nabla \cdot \bm{n}\right) ^{2} + 
K_{22}\left(\bm{n}\cdot \left[ \nabla \times \bm{n} \right] \right) ^{2}+
\right. \\ \nonumber && \left.+ 
K_{33}\left[\bm{n}\times\left[\nabla\times\bm{n}\right]\right]^{2}
\right\},
\label{frank_free_energy}
\end{eqnarray}
where $K_{11}$, $K_{22}$, and $K_{33}$ are elastic constants, and the 
integration extends over the sample volume $V$. $f_{\text{s}}$ 
is the surface anchoring energy density, which we assume to 
be of the Rapini-Papoular form (\ref{rapini_potential}), and it is 
integrated over the boundary surface $S$.

In slab geometry, the director $\bm{n}$ is assumed to lie in the 
$xz$ plane and depends only on the $z$ coordinate. Then it can be 
parametrized as  $\bm{n} = (\sin\theta,0,\cos\theta)$ which transforms
eqn~(\ref{free_energy}) into a free energy per unit area
\begin{equation}
{\cal F}_{\text{el}}/S =
\frac{1}{2}K_{33}\int_{0}^{L}{f_{13}(\theta)
\left(\frac{\partial \theta}{\partial z} \right)}^2
{dz} - \frac{1}{2}W \cos^2\theta_0,
\label{free_energy_simpl}
\end{equation}
where $f_{13}(\theta)= 1 - \delta\sin^2\theta$,
$\delta = (K_{33} - K_{11})/K_{33}$, $\theta_0 = \theta(z=0)$. 

The absence of explicit $z$-dependence in the free energy 
(\ref{free_energy_simpl}) implies the first integral
\begin{equation}
{f_{13}(\theta)\left(\frac{\partial \theta}{\partial z} \right)}^2 = 
\text{const}.
\label{first_integral}
\end{equation}
Boundary conditions read 
\begin{eqnarray}
K_{33}{f_{13}(\theta_0)\left. \frac{\partial \theta}{\partial z} 
\right|_{z=0}} 
& = & \frac{1}{2}W\sin2\theta_0, \\
\theta(z=L) & = & \theta_L.
\nonumber
\end{eqnarray}
Here we assumed that the director angle at the boundary $z=L$ is fixed.

Integrating eqn~(\ref{first_integral}), together with the boundary
conditions, yields 
\begin{eqnarray}
\label{solution_elastic}
E(\theta,\delta) =
E(\theta_0,\delta)+[E(\theta_L,\delta)-E(\theta_0,\delta)]z/L, \\
{[E(\theta_L,\delta)-E(\theta_0,\delta)]}\sqrt{f_{13}(\theta_0)} = 
(L/2\lambda)\sin2\theta_0,
\nonumber
\end{eqnarray}
where $E(\theta, \delta) = \int_{0}^{\theta}\sqrt{f_{13}(x)}dx$ is the
incomplete elliptic integral of the second kind, and $\lambda = K_{33}/W$ 
is the anchoring extrapolation length.
For small angles $\theta_0$ eqn~(\ref{solution_elastic}) can be simplified
and has the form
\begin{equation}
E(\theta,\delta) =
E(\theta_L,\delta)\frac{z+\lambda}{L+\lambda}.
\label{elastic_final}
\end{equation}
Note, that for small angles $\theta_L$ and, correspondingly, for small
$\theta$, $E(\theta,\delta) = \theta$ and we have linear dependence 
of the director angle on the $z$ coordinate
\begin{equation}
\delta n_x(z) \approx \theta(z) = \theta_L\frac{z+\lambda}{L+\lambda}.
\label{el_lin}
\end{equation}

Using eqn~(\ref{elastic_final}), one can fit the director profiles 
$\theta(z)$ with $\lambda = K_{33}/W$ and $\delta = (K_{33} - K_{11})/K_{33}$ 
as adjustable parameters. To simplify the procedure, it is more 
appropriate to fit $z(\theta)$ rather than $\theta(z)$.

\section{Onsager approach}
\label{sec:onsager}
The Helmholtz free energy in the Onsager approach is expressed in terms of 
the single-particle density, $\rho (\bm{1})$, where 
$(\bm{1}) = (\bm{r}_1,\bm{\Omega}_1)$ represents both position $\bm{r}_1$
and orientation $\bm{\Omega}_1$. It has the following form
\cite{onsager.l:1949.a,evans.r:1992.a}
\begin{eqnarray}
\nonumber
\beta {\cal F}[\rho]=
&&\int \rho (\bm{1})\left\{ \ln\rho (\bm{1})\Lambda ^{3}-1-
\beta \mu + \beta U(\bm{1})\right\} d(\bm{1})-
\\ &&
\frac{1}{2}\int 
f(\bm{1},\bm{2})\rho (\bm{1})\rho (\bm{2})d(\bm{1})d(\bm{2}).
\label{free_energy_onsager}
\end{eqnarray}
Here $\beta = 1/k_{\text{B}}T$, $\Lambda$ is the de Broglie wavelength, $\mu$ is 
the chemical potential, $U$ is the external potential energy 
(including the surface potential), 
and $f(\bm{1},\bm{2})$ is the Mayer $f$-function
\begin{equation}
f({\bf 1},{\bf 2}) = \exp \left[-{\cal V}({\bf 1},{\bf 2})/k_{\rm B}T\right]-1,
\end{equation}
where elongated particles interact pairwise
through the potential ${\cal V}({\bf 1},{\bf 2})$.

The equilibrium single-particle density that minimizes the 
free energy (\ref{free_energy_onsager}) is a solution of the following 
Euler-Lagrange equation
\begin{equation}
 \ln\rho (\bm{1})\Lambda ^{3}-\beta \mu + \beta U(\bm{1})- 
\int f(\bm{1},\bm{2})\rho (\bm{2})d(\bm{2}) = 0,
\label{dens_onsager}
\end{equation}
which can be obtained from the variation of the functional 
(\ref{free_energy_onsager}). 
In practice, we find it more convenient to minimize the functional 
(\ref{free_energy_onsager}) instead of solving the integral equation 
(\ref{dens_onsager}).

\subsection{Bulk problem}
In the bulk problem, the single particle density is independent of position,
$\rho(\bm{1}) = \rho(\bm{\Omega}_1)$. Then the integrals over position 
may be performed directly. To perform the integration over the angles, 
we expand the Mayer $f$-function 
and the single-particle density in spherical harmonics
\begin{equation}
f(\bm{1},\bm{2})=\sum_{\ell_{1},\ell_{2},\ell_{r}}f^{\ell_{1},\ell_{2},\ell_{r}}(r_{12})\Phi ^{\ell_{1},\ell_{2},\ell_{r}}(\bm{\Omega} _{1},\bm{\Omega} _{2},\bm{\Omega} _{r}),
\label{mayer_exp}
\end{equation}
\begin{equation}
\rho (\bm{\Omega})
=\sum _{\ell}^{\text{even}}\rho_{\ell}Y_{\ell 0}(\bm{\Omega}),
\label{dns_b_exp}
\end{equation}
where 
$\Phi^{\ell_{1},\ell_{2},\ell_{r}}(\bm{\Omega}_{1},\bm{\Omega}_{2},\bm{\Omega}_{r})$
is a rotational invariant \cite{gray.cg:1984.a}
\begin{eqnarray}
\label{rot_inv}
\Phi^{\ell_{1}, \ell_{2}, \ell_{r}} =
4\pi \sum_{m_{1},m_{2},m_{r}}
\left( \begin{array}{ccc}
\ell_{1} & \ell_{2} & \ell_{r}\\
m_{1} & m_{2} & m_{r}
\end{array}\right) 
 \nonumber \times \\  
Y_{\ell_{1}m_{1}}(\bm{\Omega} _{1})Y_{\ell_{2}m_{2}}(\bm{\Omega} _{2})
C_{\ell_{r}m_{r}}(\bm{\Omega} _{r}) \:.
\end{eqnarray}
Here, $\bigl(\begin{smallmatrix} \ell & \ell' & \ell'' \\ m & m' & m''
\end{smallmatrix}\bigr)$ is a $3j$ symbol, $\bm{\Omega} _{r}$ is the direction
of the unit vector $\hat{\bm{r}}_{12} = \bm{r}_{12}/r_{12}$ where $\bm{r}_{12}=
\bm{r}_{1}-\bm{r}_{2}$, $C_{\ell m}$ is a reduced spherical harmonic. In writing
eqn~(\ref{dns_b_exp}), we assumed that the director $\bm{n}_0$ is pointing along the $z$ axis.

To minimize the grand potential it is convenient to expand the logarithm 
of the density in spherical harmonics 
\begin{equation}
\ln \rho(\bm{\Omega}) = \sum _{\ell}^{\text{even}} c_\ell Y_{\ell 0}(\bm{\Omega}).
\label{lnd_b_exp}
\end{equation}
The grand potential then can be rewritten in the form
\begin{eqnarray}
\frac{\beta {\cal F}}{V} =
-\sqrt{4\pi}(1+\beta\mu)\rho_0 +  
\sum _{\ell=0}^{\text{even}} c_\ell\rho_\ell +
\frac{2\pi V_{\ell\ell}}{\sqrt{2\ell+1}}{\rho_\ell}{\rho_\ell},
\end{eqnarray}
where the coefficients $\rho_\ell$ may be expressed in terms of the 
parameters $c_\ell$, and where the pair-excluded volume is expanded in coefficients
\begin{equation}
V_{\ell\ell} = -4\pi\int_{0}^{\infty}r^2drf^{\ell\ell0}(r).
\end{equation}
The grand potential was numerically minimized with respect to variation of the
parameters $c_\ell$, by the conjugate gradient method \cite{press.wh:1992.a}.
The resulting single-particle density was used to calculate elastic constants,
for different values of the order parameter.

\subsection{Elastic constants}
To evaluate the elastic constants we used the expressions derived by 
Poniewierski and Stecki 
\cite{poniewierski.a:1979.a,lipkin.md:1985.a,yokoyama.h:1997.a} 
\begin{eqnarray}
K_{11} = M_{xxxx} = M_{yyyy}, \\
\nonumber
K_{22} = M_{xxyy} = M_{yyxx}, \\
\nonumber
K_{33} = M_{zzxx} = M_{zzyy},
\end{eqnarray}
where
\begin{eqnarray}
\label{elastic_int}
M_{\alpha\beta\gamma\delta} &=& \frac{1}{2}k_BT 
           \int d \bm{r}_{12} d \bm{\Omega}_1 d \bm{\Omega}_2 \,
           f(\bm{r}_{12}, \bm{\Omega}_1, \bm{\Omega}_2) 
           \nonumber \times \\           
            && r_{12}^{\alpha} r_{12}^{\beta} 
           \rho' (\cos\theta_1) \rho' (\cos\theta_2)
           \Omega_{1\gamma} \Omega_{2\delta}. 
\end{eqnarray}
All integrals are evaluated in a local coordinate frame with the $z$ axis
parallel to the director at point $\bm{r}$.

As discussed in appendix~\ref{sec:appendix}, performing the integrations over
the angles and making use of the properties of $3j$ symbols and spherical
harmonics, one obtains a simplified expression previously given by Stelzer {\it
  et al} \cite{stelzer.j:1995.a}:
\begin{eqnarray}
\label{elastic_constants}
\beta K_{11} &=& K(1,1,3), \\
\nonumber
\beta K_{22} &=& K(1,-5,-1), \\
\nonumber
\beta K_{33} &=& K(-2,4,-4),
\end{eqnarray}
where
\begin{widetext}
\begin{align*}
K(a_1,a_2,a_3) &=
 \frac{4\pi^2}{3} \sum_\ell^{\text{even}} 
\ell(\ell+1)\rho_\ell^2
\left[
\frac{I^{\ell,\ell,0}}{\sqrt{2\ell+1}}-
\frac{6a_1+a_2\ell(\ell+1)}{5}\sqrt{\frac{(2\ell-2)!}{(2\ell+3)!}}I^{\ell,\ell,2}
\right] +
 \\  &
 \frac{a_3}{5}\sqrt{ \frac{3}{2} }
\frac{(\ell+3)!}{(\ell-1)!}
\sqrt{\frac{(2\ell)!}{(2\ell+5)!}}
\rho_\ell\rho_{\ell+2} I^{\ell,\ell+2,2} \;,
\end{align*}
\end{widetext}
and where $I^{\ell_1,\ell_2,\ell}$ are radial integrals over the expansion 
coefficients of the Mayer $f$-function.

\subsection{Slab geometry}
In slab geometry, with the $z$ direction normal to the surfaces, the 
single-particle density and the external potential depend on the $z$ 
coordinate only. 
To perform the integrations over the angles, we expand the Mayer 
$f$-function into rotational invariants, similar to the bulk system
(\ref{mayer_exp}). The single-particle density and its logarithm can be
also expanded in spherical harmonics
\begin{eqnarray}
\nonumber
\rho (z,\bm{\Omega})
=\sum_{\ell m}\rho_{\ell m}(z)Y_{\ell m}^{*}(\bm{\Omega}), \\
\ln \rho (z,\bm{\Omega})
=\sum_{\ell m}c_{\ell m}(z)Y_{\ell m}(\bm{\Omega}). 
\label{dns_s_exp}
\end{eqnarray}
The difference from the bulk case, eqns~(\ref{dns_b_exp},\ref{lnd_b_exp}),
is that the director is allowed to vary in the $xz$ plane, so
an expansion in Legendre polynomials ($m=0$) is not sufficient.
Conducting angular and $x$- and $y$-integrations gives
\begin{eqnarray}
\lefteqn{ \frac{\beta {\cal F}}{S} =
-\int dz \sqrt{4\pi}(1+\beta\mu)\rho_{00}(z) + }  
\nonumber \\ &
\sum _{\ell ,m} \bigl(c_{\ell m}(z) + \beta U_{\ell m}(z)\bigr)\rho_{\ell m}+
\nonumber \\ &
2\pi \int dz_1 dz_2 A_{\ell_1, \ell_2, m}(z_{12})
\rho_{\ell_1 m}(z_1)\rho_{\ell_2 m}(z_2) \;,
\end{eqnarray}
where the pair-excluded area at given $z$-separation $z_{12}$ is expanded in
coefficients 
\begin{eqnarray}
A_{\ell_1,\ell_2,m}(z_{12}) = -2\pi\sum_{\ell}^{\text{even}}
\left( \begin{array}{ccc}
\ell_{1} & \ell_{2} & \ell\\
m & \overline{m} & 0
\end{array}\right) 
\nonumber \times \\ 
 \int_0^\infty s \, ds \, f_{\ell_1,\ell_2,\ell}(r_{12})P_\ell(\cos\theta_r)
\:.
\end{eqnarray}
Here $s^2 = (x_1-x_2)^2 + (y_1-y_2)^2$, $\tan\theta_r = s/(z_1-z_2)$,
$ \overline{m} = -m$.
Integrating, we took into account that the $3j$ symbol vanishes
unless $m_1+m_2+m_r=0$. 

To obtain equilibrium single-particle density profiles, the grand potential then
was tabulated on a regular grid of points $z_i$ and numerically minimized with
respect to variation of the parameters $c_{\ell m}(z_i)$, by the conjugate 
gradient method.

\subsection{Anchoring energy}
\label{sec:onsager_anch}
To obtain the microscopic expression for the extrapolation 
length $\lambda$, we start from the equation for the single-particle density
(\ref{dens_onsager}). Assume that the solution for a ground state, 
i.e. for homeotropically aligned liquid crystal in slab geometry, 
is given by the single-particle density $\rho_0$. 
Consider a small perturbation around the ground state,
$\rho = \rho_0 + \delta \rho$. To first order in $\delta \rho$, eqn 
(\ref{dens_onsager}) can be written as
\begin{equation}
\label{dens_fstord}
\frac{\delta \rho({\bf 1})} {\rho_0({\bf 1})} = 
\int f({\bf 1},{\bf 2})\delta \rho({\bf 2})d({\bf 2}).
\end{equation} 

In the case of slowly-varying director fields we  assume that the free energy 
functional is locally in equilibrium. This is equivalent to the mathematical 
simplification \cite{stelzer.j:1997.b,somoza.am:1991.a}
\begin{equation}
\rho(\bm{r}, {\bf \Omega} )=\rho_0
\left(\bm{r},\bm{n}(\bm{r})\cdot{\bf \Omega} \right).
\label{dens_perturb}
\end{equation} 
Then the density perturbation can be written in terms of the perturbation
of the director
\begin{equation}
\delta \rho(\bm{r}, {\bf \Omega} )=\rho_0 '
\left(\bm{r},\bm{n}_0\cdot{\bf \Omega} \right)
\delta \bm{n}(\bm{r})\cdot{\bf \Omega},
\label{director_perturb}
\end{equation} 
where the prime denotes a partial derivative with respect to 
$(\bm{n}_0\cdot{\bf \Omega})$.

In slab geometry, with the $z$ axis normal to the surfaces,  
the single particle density $\rho_0$ depends on the $z$ coordinate only 
and can be expanded in spherical harmonics similar to eqn~(\ref{dns_s_exp}). 
Note, that $\rho_0$ does not depend on $\phi$ which implies $m=0$ in 
expressions (\ref{dns_s_exp}). We also assume that the director is parallel to 
the $xz$ plane, $\delta \bm{n} = (\delta n_x, 0, 0)$. Conducting
angular integrations in eqn~(\ref{dens_fstord}) and making use of the 
properties of $3j$ symbols we obtain
\begin{eqnarray}
\label{int_dir}
c_{\ell_1}(z_1) \delta n_x (z_1) = 
4 \pi \sum_{\ell_2 = 2}^{\text{even}} 
\sqrt{\frac{\ell_2(\ell_2+1)}{\ell_1(\ell_1+1)}} 
\nonumber \times \\
\int_0^\infty A_{\ell_1, \ell_2, 1}(z_2 - z_1) 
\rho_{\ell_2}(z_2) \delta n_x (z_2) dz_2.
\end{eqnarray}

Equation (\ref{int_dir}) is a homogeneous Fredholm equation of the 
second kind. It allows one to calculate the director profile 
(for small director deviations from the ground state, 
$\bm{n}_0 = \bm{e}_z$) once the 
single-particle density $\rho_0$ of the ground state is known.
Eqn~(\ref{int_dir}) is not valid for every $\ell_1$, in spite of
the derivation. This is because we are trying to map the single-particle
density variation onto the director variation (eqn~\ref{dens_perturb}). 
However, this equation should be valid for the leading term in the 
density expansion, $\ell_1 = 2$, which we consider below. 

First, we construct the solution to the eqn~(\ref{int_dir}) in the cell
{\em bulk}.
The kernel $ A_{\ell_1, \ell_2, 1}(z_2 - z_1)$ is a short-ranged function. 
It equals zero for $|z_2 - z_1| > a$, where $a$ is the molecular 
length. The bulk director is a slowly-varying function on this length scale. 
Therefore we can expand it in a Taylor series
\begin{equation}
\delta n_x(z_2) = \delta n_x(z_1) + \delta n'_x(z_1)z_{12} +
                  \delta n''_x(z_1)z_{12}^2/2 + ... , 
\end{equation}
and restrict ourselves to second order in the expansion. Then 
the equation for the director (\ref{int_dir}) can be rewritten as
a second-order linear differential equation
\begin{equation}
a_2(z)\delta n''_x(z) + a_1(z)\delta n'_x(z) + a_0(z)\delta n_x(z) = 0,
\label{dir_eq}
\end{equation}
where
\begin{widetext}
\begin{eqnarray}
a_n(z) &=& \frac{4\pi}{\sqrt{6}} \sum_{\ell_2 = 2}^{\text{even}} 
\sqrt{\ell_2(\ell_2+1)} 
\int_0^L A_{2, \ell_2, 1}(z_2-z)\rho_{\ell_2}(z_2)
\frac{z_{12}^n}{n!} dz_2 
- c_2(z) \delta_{n,0}.
\end{eqnarray}
\end{widetext}

In the bulk, the expansion 
coefficients $c_{2}$ and $\rho_{\ell_2}$ do not depend on the $z$ coordinate.
Expansion coefficients of the excluded area, $A_{\ell_1, \ell_2, \ell_r}(z)$, are
even functions of $z$, which immediately implies that $a_1(z) = 0$ in 
the bulk, since the integrated function $zA_{\ell_1, \ell_2, \ell_r}(z)$ is odd 
(because $a_1(z)$ is zero in the bulk, we expanded $\delta n_x(z)$ up to
second order). 

To prove that $a_0(z) = 0$ in the bulk, we consider again the equation 
for the single particle density (\ref{dens_onsager}). Performing the
usual expansion of density and $f$-function (\ref{mayer_exp},\ref{dns_b_exp})
and integrating over the angles and $x,y$ coordinates we obtain
\begin{widetext}
\begin{eqnarray}
\nonumber
c_0(z_1) &=& \sqrt{4 \pi}\beta \mu  - 4 \pi 
\int_0^L A_{0,0,0}(z_2-z_1) \rho_0(z_2)dz_2,\\
c_{\ell_1}(z_1) &=&- 4 \pi 
 \sum_{\ell_2 = 0}^{\text{even}} \int_0^L A_{\ell_1,\ell_2,0}(z_2-z_1) 
\rho_{\ell_2}(z_2)dz_2, \hspace{0.5cm} (\ell_1 \not= 0) 
\label{bulk_dens_eq}
\end{eqnarray}
\end{widetext}
where $A_{\ell_1, \ell_2, x}$ are the expansion coefficients of the excluded area
in a series of spherical harmonics. It is easy to show that the left hand 
side of the second equation in (\ref{bulk_dens_eq}) equals $a_0(z)$ in the
cell bulk. Indeed, using the symmetry of the excluded volume 
expansion coefficients we can write
$- 4\pi\int_0^L A_{2, \ell_2, 1}(z_2-z_{\rm b})dz_2 = 
4\pi\int_{-\infty}^\infty A_{2, 2, 0}(z)dz \delta_{2,\ell_2} =
V_{22}\delta_{2,\ell_2}$, 
which converts the expression for $a_0(z)$ to the left hand side of the 
second equation in (\ref{bulk_dens_eq}). In fact, the conclusion 
$a_0(z) = 0$ in the cell bulk is a consequence of the invariance of 
the grand potential with respect to director rotations. 

Therefore, eqn~(\ref{dir_eq}) in the cell bulk reduces to $\delta n''_x(z)=0$
which means we have a linear dependence of the director on $z$ coordinate,
in agreement with the result of elastic theory (\ref{el_lin}). This also 
means that, in the cell bulk, $c_0 + c_1z$ is an eigenvector of the 
Fredholm equation (\ref{int_dir}).

To solve eqn~(\ref{int_dir}) near the cell boundary is a much more challenging
task. Numerically it can be done using, for example, an iterative method 
\cite{press.wh:1992.a}. Here we try to construct a crude analytical solution 
which will give us some qualitative understanding of what is happening in the
interface region. 

To begin with, we simplify eqn~(\ref{int_dir}). The sum over $\ell_2$ 
in the right-hand side of eqn~(\ref{int_dir}) is 
converging very fast (for ellipsoids with the elongation $e=15$, 
every term is about ten times smaller than the previous one). 
Therefore, we truncate the sum leaving only the $\ell_2=2$ term. 
Then, the kernel of the Fredholm equation $A_{2,2,1}(z_1-z_2)$ can be 
expanded in a Taylor series
\begin{equation}
\label{sep_krnl}
A_{2,2,1}(z_1-z_2) = \sum_{n=0}^{\infty} z_1^n \Phi_n(z_2),
\end{equation}
where $\Phi_n(z)$ are some functions. A practical example
of such an expansion is given in appendix \ref{sec:hermite}.
The kernel is then separable, and the problem is reduced to the solution of 
a set of linear algebraic equations. Indeed, substituting eqn (\ref{sep_krnl})
into eqn~(\ref{int_dir}) we obtain
\begin{equation}
\label{dir_int}
\delta n_x (z) = c_2^{-1}(z) \sum_{n=0}^{\infty} b_n z^n,
\end{equation}
where the $b_n$ are some constants. Substituting (\ref{dir_int}) back into 
eqn~(\ref{int_dir}) we obtain an infinite set of linear equations for the 
coefficients $b_n$. To obtain an analytical expression, we perform 
further simplifications. First we note that, in the cell bulk, 
the director is a linear function of the $z$ coordinate. Therefore, to
a good approximation, we may retain only the first two terms  
in eqn~(\ref{int_dir}) since in the cell bulk $c_2(z) = \text{const}$. 
Using again the director profile given by elastic theory, 
eqn~(\ref{el_lin}), we obtain an expression for the extrapolation length
\begin{equation}
\label{lambda_approx}
\lambda = \frac{b_0}{b_1} =
\frac{4\pi \int_0^\infty zA_{2,2,1}(z) \rho_2(z)c_2^{-1}(z) dz}
{1 - 4\pi \int_0^\infty A_{2,2,1}(z) \rho_2(z)c_2^{-1}(z)dz }. 
\end{equation}
This expression is able to give qualitative explanations of the anchoring 
phenomenon in our system. 
In the ideal case, often considered in phenomenological
approaches \cite{yokoyama.h:1997.a}, the density and order parameter are
assumed to be constant in the cell, i.e. $\rho_2(z)c_2^{-1}(z) = {\rm const}$.
The anchoring appears only due to the presence of the interface,
which breaks translational symmetry. According to eqn~(\ref{bulk_dens_eq}),
$c_2 = - V_{22} \rho_2$, therefore
$4\pi \int_0^\infty A_{2,2,1}(z) \rho_2(z)c_2^{-1}  dz = 1/2$.
Then the anchoring coefficient is proportional
to the first moment of the excluded area coefficient $A_{2,2,1}(z)$
\begin{equation}
\label{lambda_ideal}
\lambda =
- \frac{8\pi}{V_{22}} \int_0^\infty zA_{2,2,1}(z) dz,
\end{equation}
and does not depend on the value of density or order parameter, since 
excluded area, as well as excluded volume, are completely defined
by the geometry of the overlapping molecules. This means that in the ideal 
case of a uniform nematic, the anchoring coefficient $W = K_{33} / \lambda$ 
has the same dependence on the order parameter as the elastic 
constant $K_{33}$.

In reality, one observes rather strong subsurface variations of the 
density and order parameter. The ratio $\rho_2(z)/c_2(z)$ then comes into play 
and contributes to the overall dependence of the extrapolation length on the 
order parameter. The numerical results which show this dependence are 
presented in sec.~\ref{sec:onsager_slab}. A simple physical explanation
is also possible. Presmectic ordering and higher density of the nematic phase 
in the interface region indicate that the nematic liquid crystal is at 
different state point. The mesophase is more ordered at this state point and 
this ordering is less sensitive to the density variation in the cell bulk. 
Since this ordering defines the director profile at the interface, 
it also affects the dependence of the extrapolation length on the state point.

\section{Thermal fluctuations}
\label{sec:fluctuations}
Another simulation method to measure bulk elastic constants $K_{ii}$ and the 
zenithal anchoring energy $W$ is based on the measurement of the 
director fluctuation amplitudes $\delta \bm{n}$ in the liquid crystal cell 
\cite{allen.mp:1988.a,allen.mp:1990.a,allen.mp:1999.a,andrienko.d:2000.d}.
Consider again slab geometry with homeotropic anchoring of the
director at both cell surfaces. Consider a small perturbation of 
the director around the equilibrium distribution 
\begin{equation}
\bm{n}(\bm{r}) = 
\bm{e}_z + \delta\bm{n}(\bm{r}). 
\end{equation}
Minimizing the free energy (\ref{free_energy}) and linearizing the equations
for the director and boundary conditions with respect to $\delta\bm{n}$, 
we obtain
\begin{widetext}
\begin{eqnarray}
  \nonumber
  K_{11}(\delta n_{x,xx} + \delta n_{y,yx}) +
  K_{22}(\delta n_{x,yy} - \delta n_{y,yx}) + 
  K_{33}\delta n_{x,zz} = 0, 
 \\ \nonumber
  K_{11}(\delta n_{x,xy} + \delta n_{y,yy}) +
  K_{22}(\delta n_{x,xx} - \delta n_{y,yx}) + 
  K_{33}\delta n_{x,zz} = 0,
\end{eqnarray}
\begin{equation}
\left.
W\delta \bm{n}+ K_{33}\frac{\partial }{\partial z}\delta \bm{n}
\right|_{z=L} = \bm{0},
\qquad
\left.
W\delta \bm{n}- K_{33}\frac{\partial }{\partial z}\delta \bm{n}
\right|_{z=0} = \bm{0}.
\label{eqn:boundary}
\end{equation}
\end{widetext}
Here we adopt the notation $\delta n_{\alpha, \beta \gamma} = 
\partial_\beta \partial_\gamma (\delta n_{\alpha})$.

We now expand $\delta \bm{n}(\bm{r})$ in a series of orthogonal functions
\begin{align}
\lefteqn{\delta \bm{n}( \bm{r})=
\frac{1}{V}\sum_{\bm{q}_{\bot},q_{z}}
\mathrm{e}^{\mathrm{i}\bm{q}_{\bot}\cdot\bm{r}_{\bot}}
\nonumber \times} \\
&&\left[
\delta \bm{n}^{(+)}(\bm{q}_{\bot},q_{z})\mathrm{e}^{\mathrm{i}q_{z}r_{z}}+
\delta \bm{n}^{(-)}(\bm{q}_{\bot},q_{z})\mathrm{e}^{-\mathrm{i}q_{z}r_{z}}
\right],
\label{eqn:fluct}
\end{align}
where $\bm{q}_{\bot}=(q_{x},q_{y})$, and
$\delta \bm{n}^{(-)}=(\mathrm{i}\chi-\xi)/(\mathrm{i}\chi+\xi)
\delta \bm{n}^{(+)}$ to satisfy the boundary conditions.
Here we have introduced the dimensionless wave vector $\chi =q_{z}L$
and anchoring parameter
\begin{equation}
\xi =\frac{WL}{K_{33}}=\frac{L}{\lambda } \:,
\label{eqn:anchoring}
\end{equation}
where $\lambda$ is the extrapolation length
\cite{degennes.pg:1995.a}.
The wave vectors $q_{z}$ form a discrete spectrum
because of confinement in the $z$ direction
which depends on the anchoring energy $W$.
The explicit form of the $q_{z}$ spectrum is given by the secular equation:
\begin{equation}
\left(\xi^{2}-\chi^{2}\right) \sin\chi +2\xi\chi \cos\chi =0 \:.
\label{eqn:secular}
\end{equation}

In molecular simulations, rather than measuring director fluctuations,
it is more convenient to measure fluctuations of the
second-rank order tensor components
(following Ref.~\cite{forster.d:1975.a}).
We define the real-space order tensor density
\begin{eqnarray*}
Q_{\alpha\beta}(\bm{r}) &=&\frac{V}{N}\sum_{i}
\delta(\bm{r}-\bm{r}_{i}) Q_{\alpha\beta }^{i}
\:, \\
Q_{\alpha\beta}^{i} &=&
\frac{3}{2}\left(u_{i\alpha}u_{i\beta}-\frac{1}{3}\delta_{\alpha\beta}\right)
\:,
\end{eqnarray*}
where $\alpha ,\beta =x,y,z$,
in terms of the orientation vectors
$\bm{u}_{i}$ of each molecule $i$ (we consider only uniaxial molecules).
If we assume that there is no variation in the degree of ordering, and neglect
biaxiality of the order tensor,
we may write
\[
Q_{\alpha\beta}(\bm{r}) =
\frac{3}{2}Q n_{\alpha}(\bm{r})n_{\beta}(\bm{r})
- \frac{1}{2}Q\delta_{\alpha\beta},
\]
where $Q$ is the order parameter,
i.e.\ the largest eigenvalue of $Q_{\alpha\beta}(\bm{r})$.

Measurements are performed directly in reciprocal space.
The Fourier transform of the real-space order tensor is
\[
Q_{\alpha\beta}(\bm{k}) =
\int_V ~ Q_{\alpha\beta}(\bm{r}) \mathrm{e}^{\mathrm{i}\bm{k}\cdot\bm{r}}d\bm{r}
=
\frac{V}{N}\sum_i~Q_{\alpha\beta}^{i} \mathrm{e}^{\mathrm{i}\bm{k}\cdot\bm{r}_i}
\:.
\]
Then the fluctuations
$\left\langle\left| Q_{\alpha\beta}(\bm{k})\right|^{2}\right\rangle$
can be easily measured from simulations
\begin{eqnarray}
 \nonumber
\left| Q_{\alpha\beta}({k_z})\right| ^{2}
=\frac{V^{2}}{N^{2}}
\left[\left(\sum_i~Q_{\alpha\beta}^i\cos({k_z}\cdot{z}_i)\right)^2 + 
 \right. \\ \left. 
\left(\sum_i~Q_{\alpha\beta}^i\sin({k_z}\cdot{z}_i)\right)^2
\right] \:,  
\label{eqn:simul}
\end{eqnarray}
and compared with the prediction of elastic theory
\begin{eqnarray}
 \nonumber
\left\langle \left| Q_{\alpha z}(k_{z}) \right|^{2}\right\rangle 
=
\frac{9}{8}k_{\text{B}}T\frac{Q^{2}V}{K_{33}}\sum_{q_{z}}~
\frac{\chi^2+\xi^2}{q_{z}^{2}\left( 2\xi +\chi ^{2}+\xi ^{2}\right) }
\times \\  
\left| 
\frac{\mathrm{e}^{\mathrm{i}(\kappa+\chi)}-1}{\kappa+\chi}
+
\left(\frac{\mathrm{i}\chi -\xi }{\mathrm{i}\chi +\xi }\right)
\frac{\mathrm{e}^{\mathrm{i}(\kappa-\chi)}-1}{\kappa-\chi}
\right|^{2}
\label{eqn:theory}
\end{eqnarray}
where $\left\langle\ldots\right\rangle$ denotes an ensemble average,
$\kappa = k_z L$.

We measure $Q$ and
$\left\langle\left| Q_{\alpha z}( k_{z}) \right|^{2}\right\rangle$
from simulations, eqn~(\ref{eqn:simul}),
and then compare with the theoretical prediction,
eqn~(\ref{eqn:theory}),
which is parametrized by $L$, $\lambda$ and $K_{33}$.
Both the permitted $q_z$ spectrum, and the variation of
$\left\langle\left|Q_{\alpha z}(k_{z})\right|^2\right\rangle$
with $k_z $, are sensitive to the anchoring strength
parameter $\xi=L/\lambda$ .

\section{Molecular model and simulation methods}
\label{sec:simulation}
We performed Monte Carlo (MC) simulation of a liquid crystal confined 
between parallel walls, with finite homeotropic anchoring at the walls.
We used a molecular model which has been studied earlier in this geometry 
\cite{allen.mp:1999.a}. The molecules in this study 
were modeled as hard ellipsoids of revolution of elongation $e=a/b=15$, 
where $a$ is the length of the major axis 
and $b$ the length of the minor axis.
Such systems cannot form smectic or columnar phases \cite{frenkel.d:1984.a}, 
and the phase transitions are not thermally driven as they are for most 
mesogens. Therefore, it should be borne in mind that simulation results 
are model specific, however the advantage of using systems of hard ellipsoids 
is that they exhibit a simple phase behaviour with some resemblance of 
that of real nematogens. 
In addition, as the elongation increases, the nematic-isotropic phase 
boundary approaches the Onsager limit. This eases the comparison between 
simulation and density-functional Onsager-type theories.

The phase diagram and properties of this family of models are well studied 
\cite{frenkel.d:1984.a,frenkel.d:1985.a,tjiptomargo.b:1992.a,allen.mp:1993.g,allen.mp:1993.h}. 
It is useful to express the density as a fraction of the close-packed 
density $\rho _{\text cp}$ of perfectly aligned hard ellipsoids, assuming
an affinely-transformed face-centered cubic or hexagonal close packed lattice. 
For this model, temperature is not a significant thermodynamic quantity, 
so it is possible to choose $k_{\text{B}}T=1$ throughout.

The slab geometry is defined by two hard parallel confining walls, which
cannot be penetrated by the \emph{centers} of the ellipsoidal molecules.
Packing considerations generate homeotropic ordering at the surface, 
since the molecules have more free volume if their axes are normal to the wall. 
Surface anchoring in this system has been studied recently by applying an 
orienting perturbation at one of the walls and observing the response at 
the other \cite{allen.mp:1999.a} and by measuring
amplitudes of the director fluctuations \cite{andrienko.d:2000.d}.
This yielded an estimate of the extrapolation length
$\lambda\approx 17.5b \approx 1.16a$ at one particular state point 
corresponding to the value of the order parameter $Q \approx 0.85$.

A sequence of runs was carried out for systems of $N=2000$ particles using 
the constant-$NVT$ ensemble, allowing typically $10^{5}$ MC sweeps for 
equilibration and $4\times10^{7}$ sweeps for accumulation of averages
(one sweep is one attempted move per particle).
The wall separation was fixed and equal to $ L_z \approx 4.93a$. Note, 
that the size of the simulation box $L_z$ is not equal to the liquid 
crystal cell thickness $L$ appearing in the elastic theory. The difference
can be ascribed to partial penetration of the walls by the liquid crystal 
molecules, and/or formation of a solid layer near the walls. 
It is possible to write $L = L_z + 2L_w$, where $L_w$ characterizes
the wall. In our previous publication we found $L_w \approx 0.3a$ 
\cite{andrienko.d:2000.d}.

The same molecular model and interaction of the molecules with the walls was
adopted for the Onsager calculations. It was found sufficient to include terms
with $0 \le l \le 8$ in the expansion of $\ln \rho$, while terms with $0 \le l
\le 10$ were included in the expansion of the pair excluded volume coefficients
$V_{\ell\ell}$ and the excluded area coefficients $A_{\ell_1,\ell_2,m}$.

\section{Results and discussion}
\label{sec:results}
\subsection{Onsager theory, bulk}
Minimization of the free energy functional (\ref{free_energy_onsager}) 
was carried out for several values of chemical potential $\mu$. 
From the single-particle density we evaluated Frank elastic constants, 
which, together with the values of the bulk density as a fraction of the 
close-packed density and the value of the order parameter are given in 
table~\ref{tab:1}.
\begin{table}
\caption[Parameters]{
\label{tab:1}
Onsager calculations in the bulk. Reduced densities $\rho / \rho_{\text{cp}}$, 
values of the nematic order parameter $Q$, 
and elastic constants for hard ellipsoids with elongation $e = 15$.} 
\begin{ruledtabular}
\begin{tabular}{dddddd}
\multicolumn{1}{c}{$\mu$} & 
\multicolumn{1}{c}{$\rho / \rho_{\text{cp}}$} & 
\multicolumn{1}{c}{$Q$} & 
\multicolumn{1}{c}{$K_{11}$} & 
\multicolumn{1}{c}{$K_{22}$} & 
\multicolumn{1}{c}{$K_{33}$}  \\
\hline
 1.2 & 0.0293 & 0.7193 & 0.7951 & 0.2925 & 3.5290 \\
 1.3 & 0.0303 & 0.7532 & 0.8883 & 0.3294 & 4.4024 \\
 1.4 & 0.0313 & 0.7779 & 0.9677 & 0.3614 & 5.2387 \\
 1.5 & 0.0323 & 0.7973 & 1.0398 & 0.3909 & 6.0704 \\
 1.6 & 0.0332 & 0.8132 & 1.1076 & 0.4190 & 6.9113 \\
 1.7 & 0.0341 & 0.8266 & 1.1726 & 0.4463 & 7.7694 \\
 1.8 & 0.0349 & 0.8382 & 1.2361 & 0.4731 & 8.6496 \\
 1.9 & 0.0358 & 0.8482 & 1.2985 & 0.4997 & 9.5547 \\
 2.0 & 0.0367 & 0.8571 & 1.3605 & 0.5263 & 10.4883 \\
 2.1 & 0.0376 & 0.8650 & 1.4225 & 0.5531 & 11.4510 \\
 2.2 & 0.0384 & 0.8721 & 1.4848 & 0.5801 & 12.4448 \\
 2.3 & 0.0393 & 0.8785 & 1.5477 & 0.6076 & 13.4716 \\
 2.4 & 0.0402 & 0.8843 & 1.6113 & 0.6354 & 14.5318 \\
 2.5 & 0.0410 & 0.8896 & 1.6758 & 0.6639 & 15.6268 
\end{tabular}
\end{ruledtabular}
\end{table}
The results are typical for the elastic constants of prolate bodies.
They increase with fluid density (order parameter) and $K_3 > K_1 > K_2$
\cite{tjiptomargo.b:1992.a}.

It is essential to carry out bulk calculations if we want to know both anchoring
extrapolation length $\lambda = K_{33}/W$ and anchoring strength $W$.  The bulk
problem provides us with the absolute values of the elastic constants; the
elastic theory includes only ratios of elastic constants in the expression for
the director profile.

\subsection{Onsager theory, slab geometry}
\label{sec:onsager_slab}
Minimization of the grand potential was carried out in slab geometry for the
same values of chemical potential $\mu$ as considered in bulk. From the density
and order parameter profiles we were able to extract values of these quantities
in the central part of the cell which agreed with the bulk Onsager calculations.

\begin{figure}
\includegraphics[width=8cm]{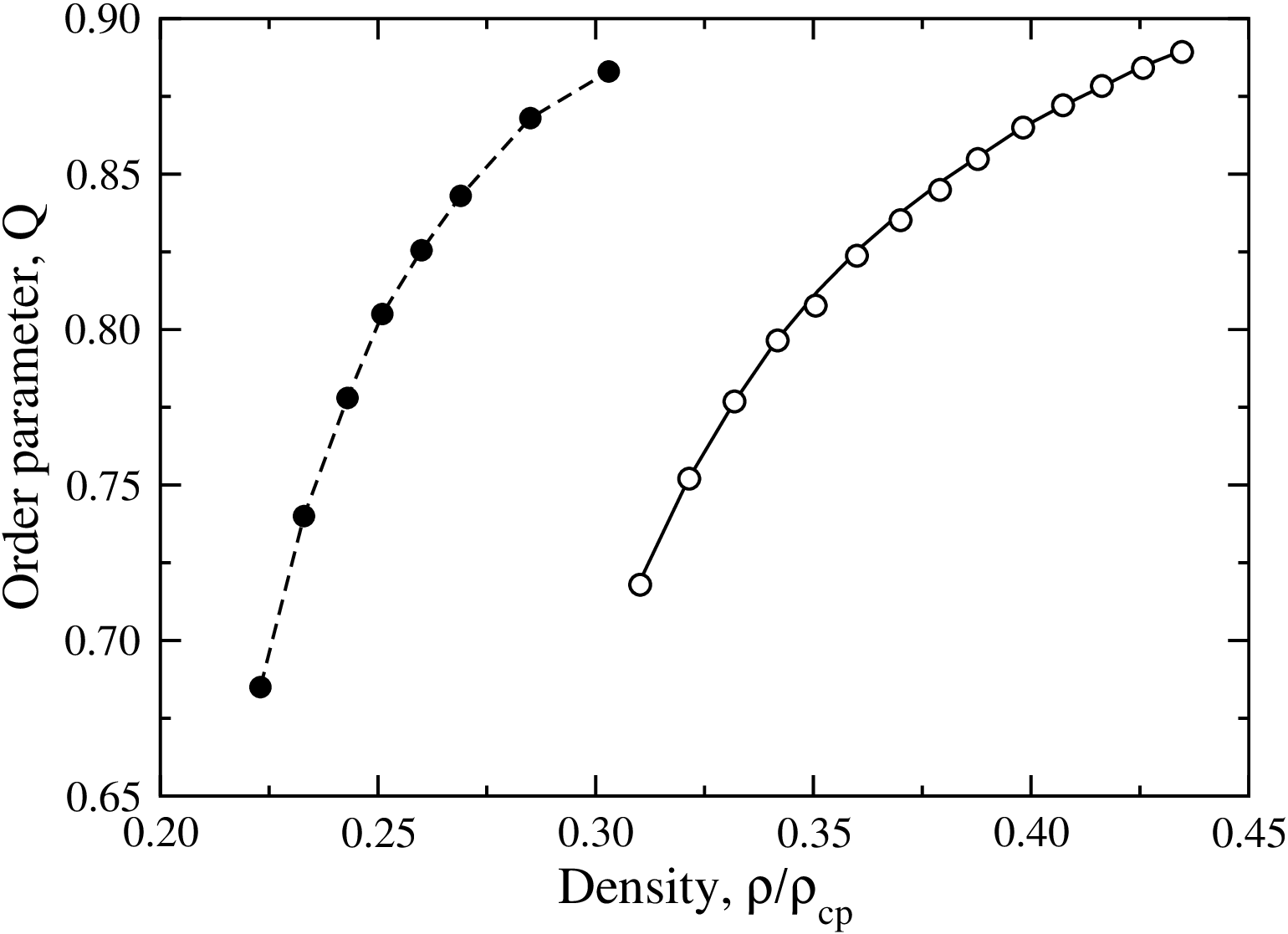}
\caption[]{
  Order parameter $Q$ as a function of the density (measured far from the
  surface).  The density is expressed relative to the closed-packed density
  $\rho_{\text{cp}}$ for perfectly aligned ellipsoids. Filled circles: simulation
  results. Open circles: Onsager theory calculations in slab geometry. Solid
  line: Onsager theory, bulk.  }
\label{fig:order_density} 
\end{figure}

Together with the simulation results, we plot dependences of the bulk order 
parameter versus density in Fig.~\ref{fig:order_density}. 
The Onsager theory does not describe the bulk equation for the state 
perfectly: the predicted bulk density for a given order parameter 
value is larger than the value obtained in the simulation.

\begin{figure}
\includegraphics[width=8cm]{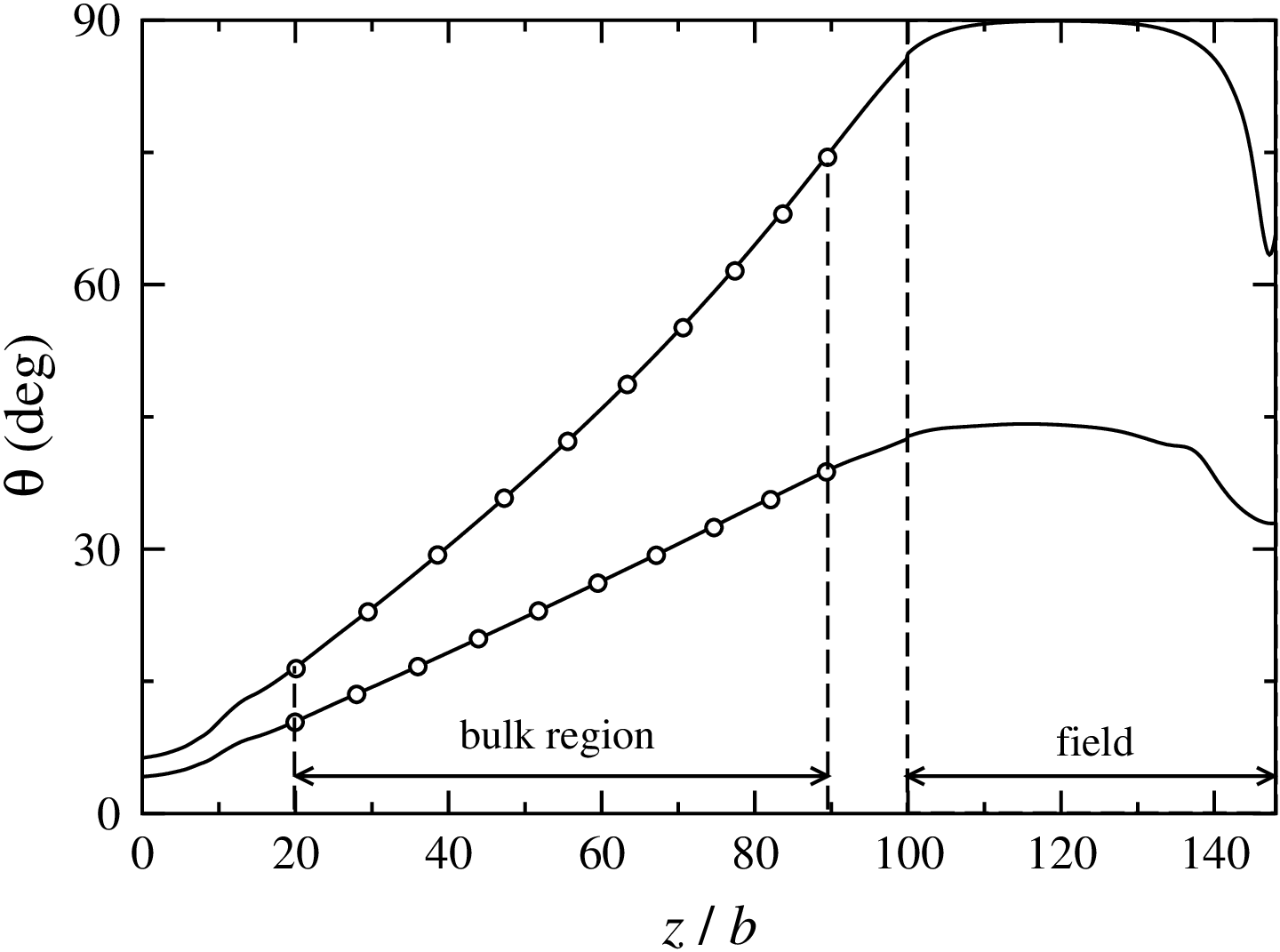}
\caption[]{
  Typical profiles of director angle $\theta(z)$ in the slab geometry. Orienting
  fields are applied in the region $100 < z/b < 148$ near the right wall,
  favouring director angles of $\pi/4$ and $\pi/2$ relative to the surface
  normal.  The left wall is unperturbed.  Solids lines: Onsager theory. Circles:
  results of fitting the profiles in the bulk region with the prediction of
  elastic theory.  }
\label{fig:director_onsager}
\end{figure}

With this choice, the same values of the chemical potential and the same 
slab dimensions, minimization of the grand potential was carried out for 
the system with an external field applied near the right wall. 
Profiles of director angle $\theta(z)$ are compared with elastic theory 
(eqn~(\ref{elastic_final})) in Fig.~\ref{fig:director_onsager}.  
The elastic theory has been fitted to the director angle profiles 
predicted by the Onsager theory using two adjustable parameters,
the extrapolation length $\lambda$ and the elastic constant ratio $\delta$.
Note that only part of the bulk region has been used for fitting, 
$ 20 < z/b < 80$ where the elastic theory is applicable. 

\begin{figure}
\includegraphics[width=8cm]{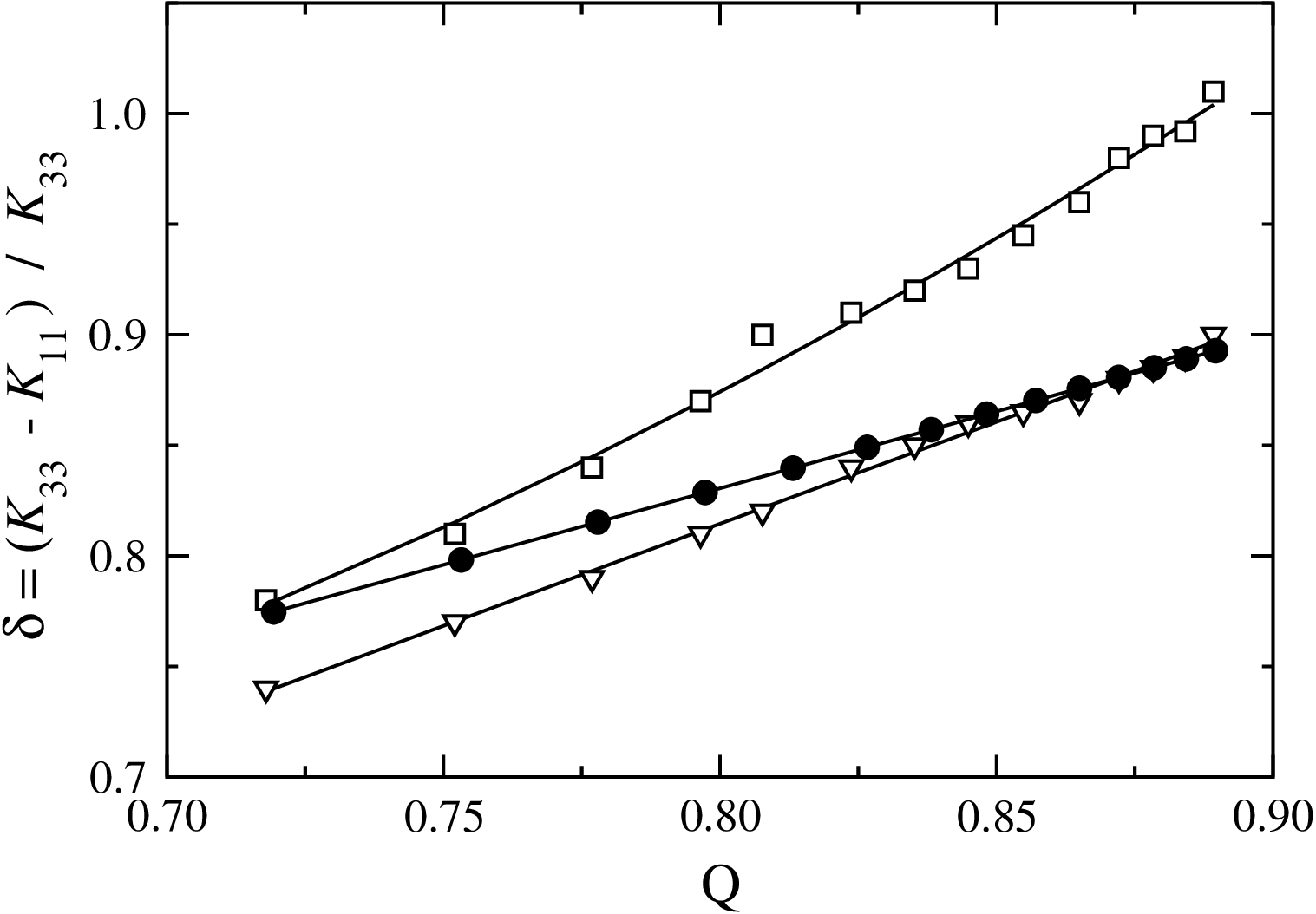}
 \caption[]{
   Ratio of the elastic constants, $\delta = (K_{33} - K_{11})/K_{33}$. Circles:
   Poniewierski - Stecki expressions. Squares: Onsager theory in slab geometry,
   with wall anchoring field at angle $\alpha=\pi/4$, fitted with the results of
   elastic theory. Triangles: Onsager theory with $\alpha=\pi/2$, fitted with
   elastic theory.  }
\label{fig:delta} 
\end{figure}

The anisotropy of elastic constants $\delta$ obtained from fitting is 
shown in Fig.~\ref{fig:delta}, together with the results of 
calculations using the Poniewierski and Stecki expressions.
The value of $\delta$, and hence the splay constant $K_1$, comes into play only
as the deformation angle $\theta(z)$ increases.  Therefore, for the external
field with easy axis $\theta_L=\pi/4$, the error in the determination of
$\delta$ from simulation data is quite large.

\begin{figure}
\includegraphics[width=8cm]{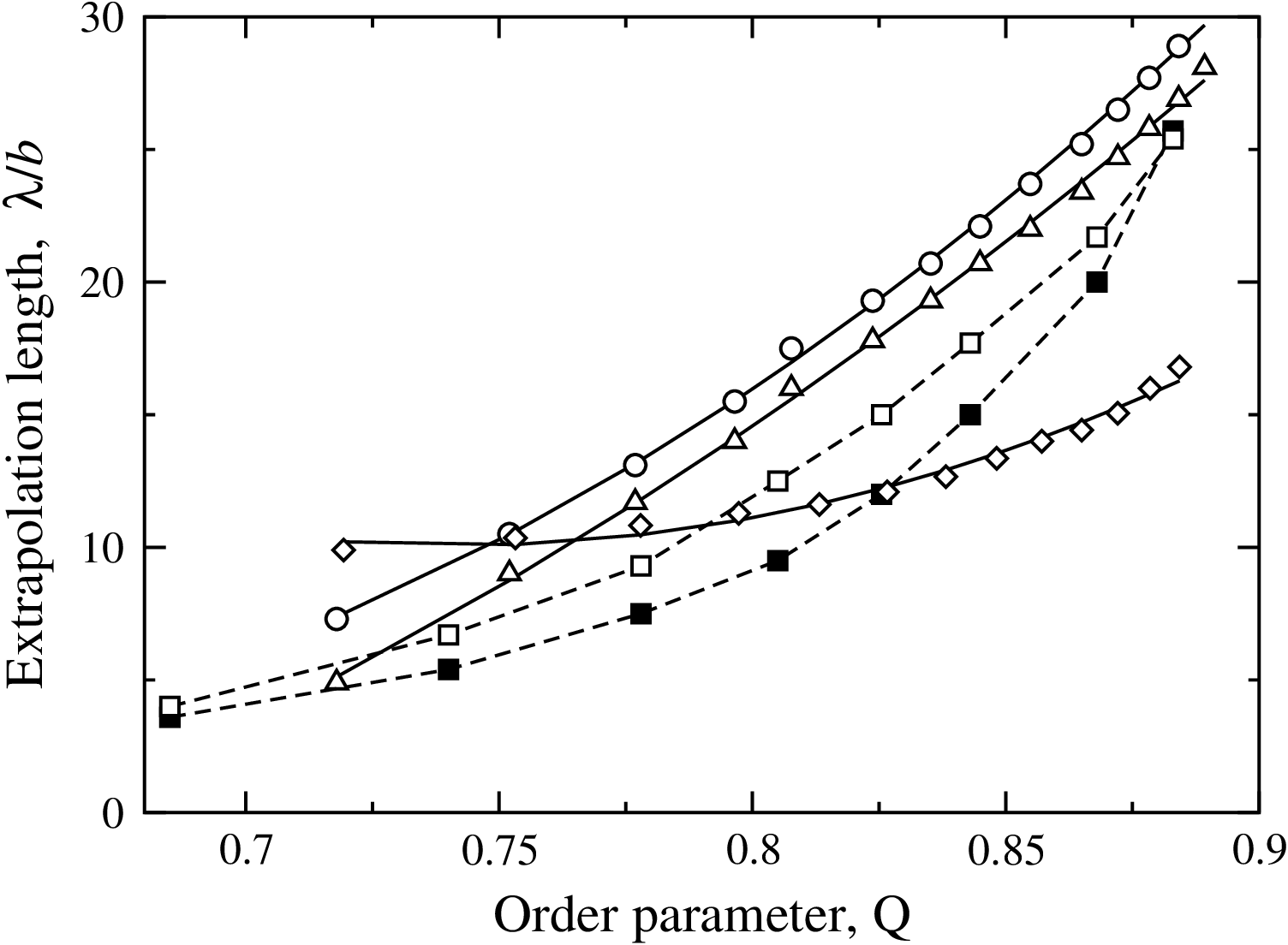}
 \caption[]{
   Extrapolation length $\lambda$ as a function of the order parameter.
   Circles: Onsager theory in slab geometry, director profiles fitted with
   elastic theory, for anchoring field with $\alpha=\pi/4$.  Triangles: the
   same, but for $\alpha=\pi/2$.  Diamonds: Onsager theory in slab geometry,
   with no field, extrapolation length calculated using equation
   (\ref{lambda_approx}).  Filled squares: Monte Carlo results obtained by
   measuring director fluctuations.  Open squares: Monte Carlo results, with
   applied field near the right-hand wall.  }
\label{fig:lambda} 
\end{figure}

The dependence of the extrapolation length, $\lambda$, on the order 
parameter $Q$ is shown in Fig~\ref{fig:lambda}. It is seen that Onsager
theory predicts the extrapolation length to grow linearly with the order 
parameter, which is completely different from the experimental results 
in thermotropic liquid crystals, where $\lambda$
decreases with the increase of the order parameter as $Q^{-2}$. 

We have also carried out minimization of the grand potential for the 
system without an external field. As a result, we obtained the single-particle 
density with a homeotropic distribution of the director in 
the cell. 
The dependence of the extrapolation length on the order parameter
was then calculated using eqn~(\ref{lambda_approx}) and is also shown 
in Fig~\ref{fig:lambda}.
The results qualitatively agree with the results of fitting obtained by
combining Onsager theory and elastic theory. The extrapolation length
tends to grow with increase in the order parameter and has the same order
of magnitude.  
Plugging this single-particle density into all equations preceding 
 eqn~(\ref{lambda_approx}) we were able to check the approximations we 
did deriving this equation. We found that the most crude approximation is the
truncation of the sum in eqn~(\ref{dir_int}). This is not a fundamental
problem and can be easily corrected by taking into account a sufficient 
number of expansion coefficients. However, this points out that the dependence 
of the director on the $z$ coordinate near the cell surface is different from 
the linear dependence in the cell bulk.
Moreover, to obtain correct quantitative values of the anchoring coefficient, 
or extrapolation length, one needs to know the director distribution 
at the surface. 
The assumption $\delta \bm{n} = \text{const}$ in the interface region, which has 
been made in \cite{stelzer.j:1997.b}, may lead to absolutely 
incorrect estimates of the anchoring coefficient.
\subsection{Simulation}
Simulations were carried out in slab geometry for several values 
of the number density. The density variation 
$0.25\leq\rho/\rho_{\text{cp}}\leq0.34$ provides a sufficient range of order 
parameter variation in the nematic phase, $0.68\leq Q\leq0.88$ for us to study
the effect of ordering on elastic behaviour. 

The order tensor fluctuations in reciprocal space were calculated 
using expression (\ref{eqn:simul}). To fit the simulation results
we used expression (\ref{eqn:theory}). Recall that the
size of the simulation box $L_z$ is not necessarily equal to the 
liquid crystal cell thickness $L$ appearing in the elastic theory. 
We found that $L=L_z+2L_{\text{w}}$, with $L_{\text{w}} \approx 4.5b = 0.3a$, 
almost independent of the density.

\begin{figure}
\includegraphics[width=8cm]{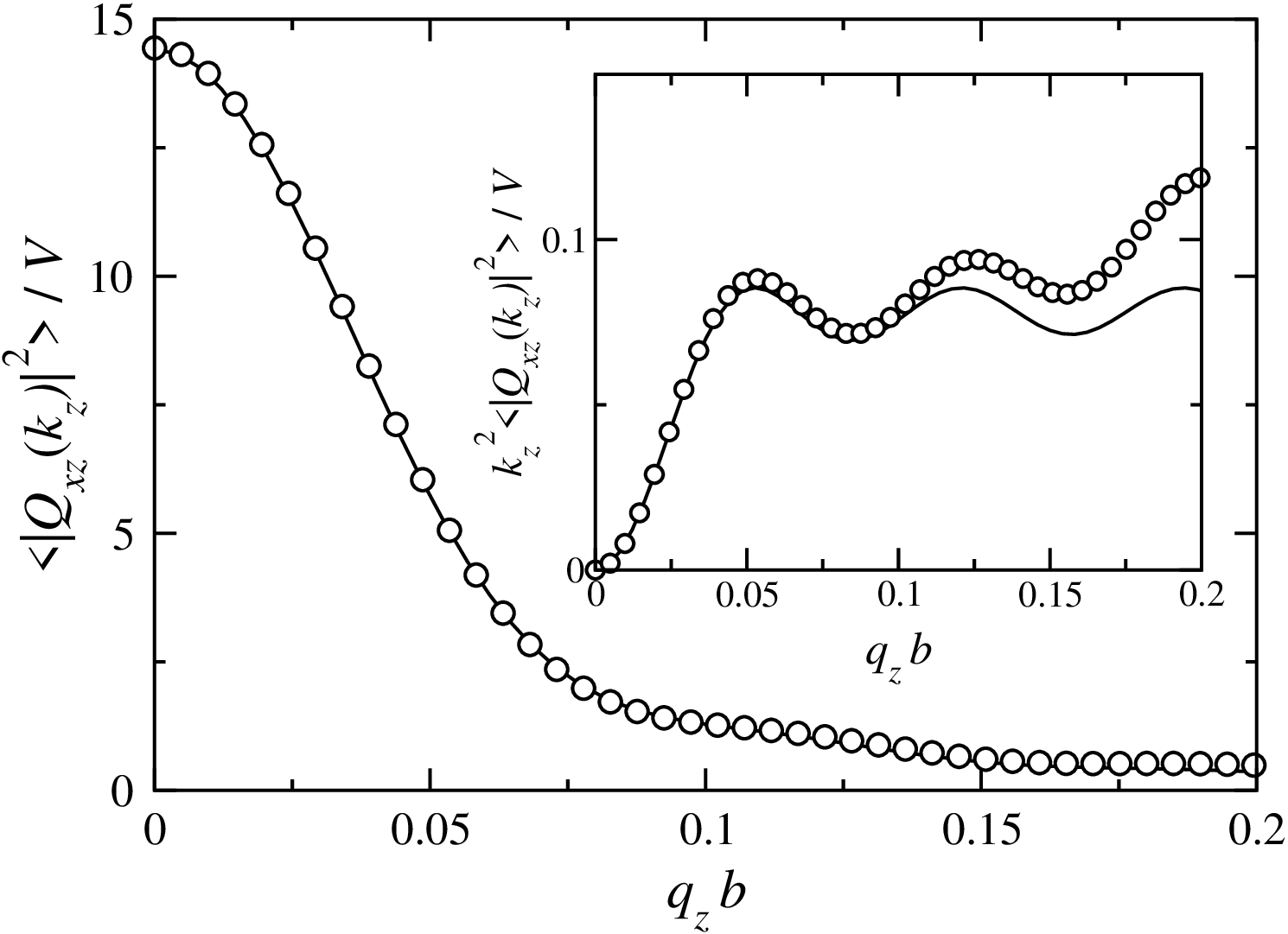}
\caption[]{
Fluctuations of the director (arbitrary units) as a function of wave-vector 
(normalized by the molecular minor axis length $b$). 
Symbols: Monte Carlo results. Solid lines: elastic
theory, fitted to parameters discussed in the text. Inset: fluctuations
multiplied by $(k_zb)^2$ to emphasize structure at higher wave numbers.
}
\label{fig:amplitudes} 
\end{figure}

Using this value of $L_{\text{w}}$ we adjusted the extrapolation length $\lambda$
and the elastic constant $K_{33}$ to obtain the best fit to the 
fluctuation amplitudes.
Typical order fluctuation amplitudes together with the corresponding 
fitting curves 
are plotted in Fig~\ref{fig:amplitudes}. 
From this plot one can see that the 
fitting curves agree quite well with the simulation results for small 
values of the wave-vector $k_z$. At higher $k_z$, the structure is not
perfectly reproduced, as one would expect for a theory valid for 
long-wavelength fluctuations, but the agreement is satisfactory. 

\begin{figure}
\includegraphics[width=8cm]{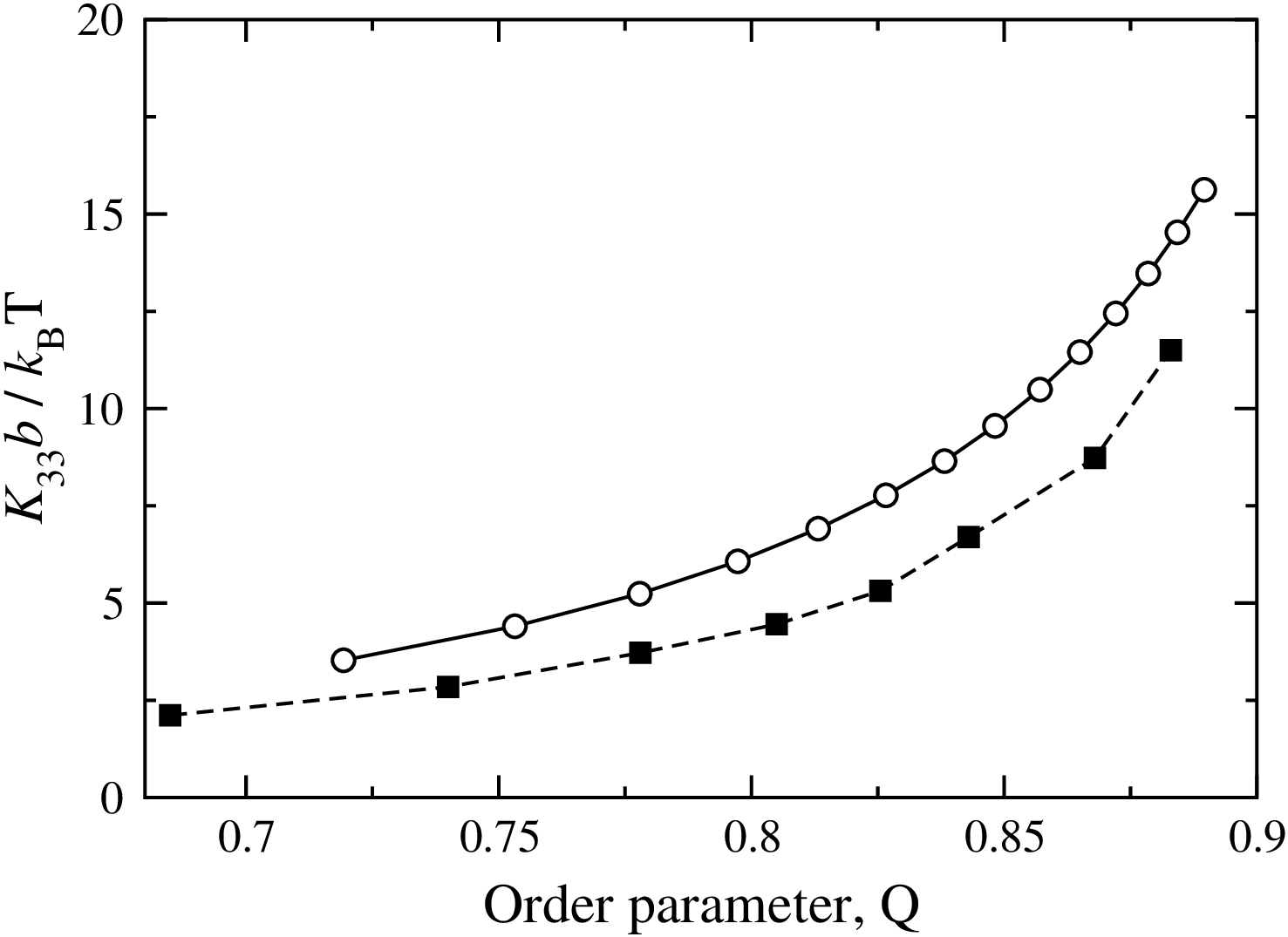}
 \caption[]{
   $K_{33}b/k_{\text{B}}T$. Circles: Onsager bulk calculations. Squares: MC
   simulation results.  }
\label{fig:k33} 
\end{figure}

The elastic constant, $K_{33}b/k_{\text{B}}T$ versus order parameter 
$Q$ is plotted in Fig~\ref{fig:k33}. The agreement 
with the Onsager theory is satisfactory, especially if we take into 
account that the equation of state is not perfectly reproduced by the 
Onsager theory.

The dependence of the extrapolation length, $\lambda$, on the order parameter
 $Q$ is shown in Fig~\ref{fig:lambda}, together with the results from the 
Onsager theory. It should be noted that combining the elastic approach with
the Onsager calculations does not allow one to determine, separately, 
$L_{\text{w}}$ and $\lambda$. Therefore, the results of the Onsager theory in 
Fig~\ref{fig:lambda} really represents $\lambda + L_{\text{w}}$, which is 
one of the possible origins of the systematic difference between 
the two approaches.

To double check the results obtained by examining the director fluctuation
amplitudes, we performed the same type of experiments as in the Onsager
slab system. Within a range $7.5b$ of the right-hand wall, a strong coupling
field was applied to molecular orientations, 
$V^{\text{ext}} \sim (\bm{u}_i \cdot \bm{e}_z)^2$, aligning the molecules
near the right wall parallel to it. After the system was equilibrated, 
the director profile was fitted to the result of the elastic theory, 
eqn~(\ref{elastic_final}). 
The dependence of the extrapolation length $\lambda$ 
on the order parameter $Q$ is also shown in 
Fig~\ref{fig:lambda}. One can see that the agreement between the two 
methods is quite good.
\begin{figure}
\includegraphics[width=8cm]{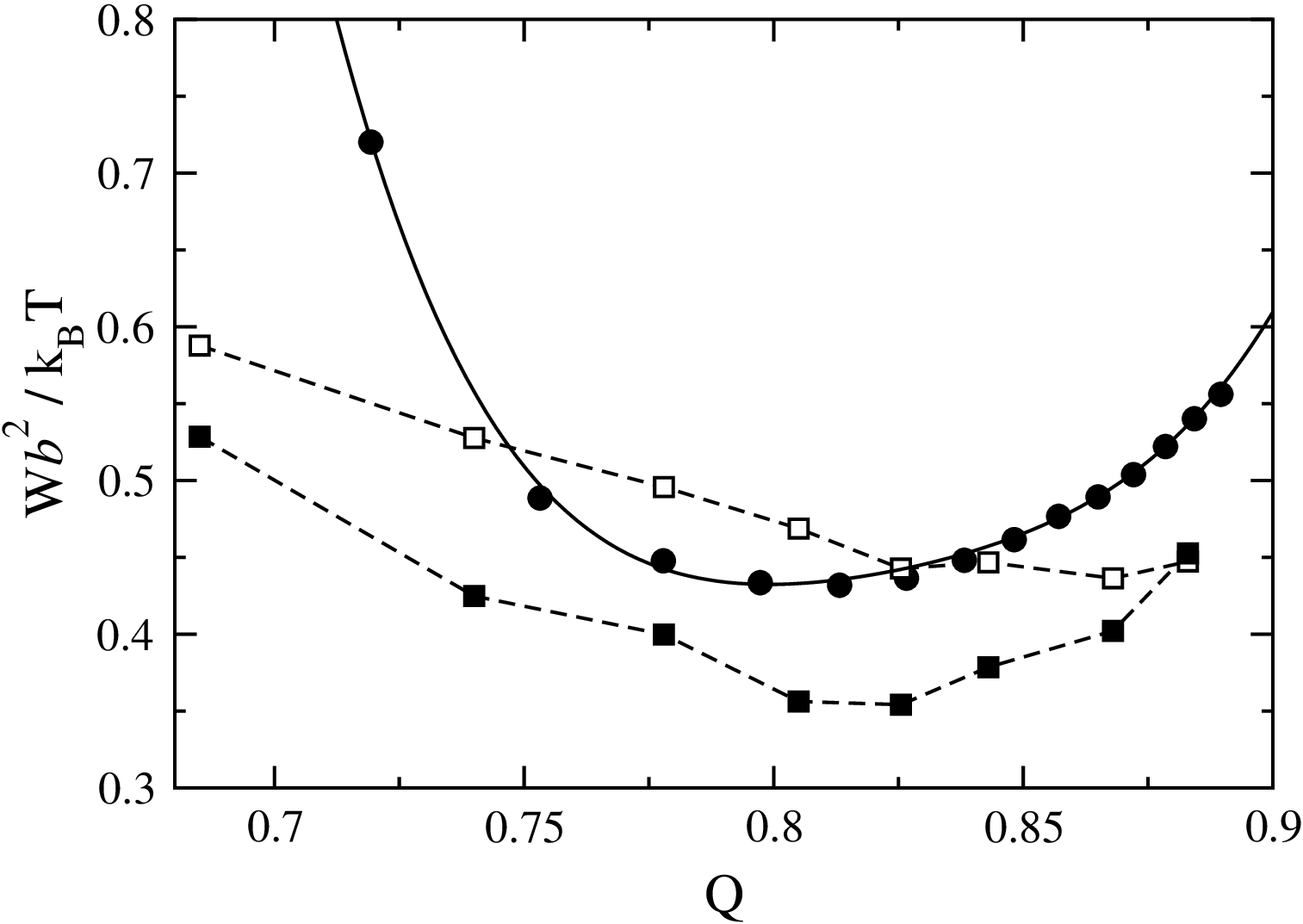}
 \caption[]{
Anchoring energy $W$ as a function of the order parameter. 
Circles: Onsager theory (with a smooth curve to guide the eye). 
Open squares: MC results obtained by measuring director fluctuations.
Filled squares: MC results, field near the right-hand wall.
}
\label{fig:W} 
\end{figure}

Finally, we plot the dependence of the anchoring energy coefficient,
$W = K_{33}/\lambda$, which is an intrinsic charachteristic of the 
interface region, in Fig.~\ref{fig:W}. For Onsager theory, 
$\lambda(Q)$ is given by the results in slab geometry
fitted with the elastic theory (Fig.~\ref{fig:lambda}), 
$K_{33}(Q)$ - by the Poniewierski - Stecki expressions (Fig.~\ref{fig:k33}). 
For MC simulations, we used the elastic constant obtained from the analysis
of the director fluctuation amplitudes. All methods predict that the
anchoring coefficient is a non-monotonic function of the order parameter,
even though the actual variation is small.

\section{Conclusions}
We have studied the dependence of the zenithal surface anchoring coefficient
on the order parameter of a lyotropic nematic liquid crystal modeled 
by hard ellipsoids.
Several techniques have been used: Onsager theory combined with elastic 
theory; Monte Carlo simulations fitted to elastic theory; 
analysis of the director fluctuation amplitudes obtained from 
Monte Carlo simulations.
The results of these methods agreed qualitatively with each other. 

Perhaps the most interesting aspect of this study is the increase of the 
anchoring extrapolation length with the increase of the nematic 
order parameter. 
This implies that the bulk elastic moduli in our system grow faster than 
the surface anchoring strength. This is opposite to the experimentally 
observed behaviour in thermotropic nematics, where the extrapolation 
length goes down with increase of the order parameter. 

A microscopic semi-qualitative expression for the extrapolation length allowed us
to conclude that subsurface variations of the single-particle density,
mainly defined by the nematic order parameter and density variation, 
contribute substantially to the anchoring phenomenon. 
We showed that for an ideal system, in which the
single-particle density in the cell is assumed to be uniform, 
the extrapolation length does not depend on the nematic order 
parameter. This dependence is therefore associated with the subsurface 
variations of the single-particle density. 

We now turn to a brief discussion about possible generalizations of this 
work. First, it would be interesting to go beyond the limits of Onsager
theory and use the direct pair correlation function in the nematic 
liquid instead of the Mayer $f$-function. Second, as was shown 
in sec.~\ref{sec:onsager_anch}, the anchoring 
coefficient can be calculated if we know the single-particle density and 
the direct pair correlation function (or Mayer $f$-function in
case of Onsager theory). This has also been done by Stelzer {\it et al} 
\cite{stelzer.j:1997.b} with numerous approximations. 
Combining elastic theory and local density functional theory one
can avoid these approximations, or at least check their 
validity. This work is in progress.

\begin{acknowledgments}
This research was supported by EPSRC grants GR/L89990 and
GR/M16023.  D.A. acknowledges the support of the Overseas Research Students 
Grant; M.P.A. is grateful to the Alexander von Humboldt foundation, the British
Council ARC programme, and the Leverhulme Trust.
\end{acknowledgments}
\appendix
\section{Elastic constants}
\label{sec:appendix}
Performing the integrations over the angles in expression (\ref{elastic_int}) 
and making use of the properties of $3j$ symbols and spherical harmonics 
one obtains
\begin{eqnarray}
\label{elastic_stelzer}
\beta K_{11} &=& \kappa(1,1,1,1,1,1), \\
\nonumber
\beta K_{22} &=& \kappa(1,-1,1,-1,1,-1), \\
\nonumber
\beta K_{33} &=& \kappa(-2,0,-2,0,-2,0),
\end{eqnarray}
where
\begin{widetext}
\begin{eqnarray}
\nonumber
\lefteqn {\kappa(b_1,b_2,b_3,b_4,b_5,b_6) =} \\
\nonumber
&& \frac{4\pi^2}{3} \sum_\ell^{\text{even}} 
\ell(\ell+1)\rho_\ell^2 
\left[
-\left(\begin{array}{ccc} \ell&\ell&0\\ 1&-1&0 \end{array}\right)I^{\ell,\ell,0}+
\frac{b_1}{5}
\left(\begin{array}{ccc} \ell&\ell&2\\ 1&-1&0 \end{array}\right)I^{\ell,\ell,2}+
\frac{b_2}{5}\sqrt{\frac{3}{2}} 
\left(\begin{array}{ccc} \ell&\ell&2\\ 1&1&-2 \end{array}\right)I^{\ell,\ell,2}
\right] + \\
\nonumber
&& \sqrt{\frac{(\ell+1)!}{(\ell-3)!}}\rho_\ell\rho_{\ell-2}
\left[
\frac{b_3}{5}
\left(\begin{array}{ccc} \ell&\ell-2&2\\ 1&-1&0 \end{array}\right)+
\frac{b_4}{5}\sqrt{\frac{3}{2}} 
\left(\begin{array}{ccc} \ell&\ell-2&2\\ 1&1&-2 \end{array}\right)
\right]I^{\ell,\ell-2,2} + \\
&& \sqrt{\frac{(\ell+3)!}{(\ell-1)!}}\rho_\ell\rho_{\ell+2} 
\left[
\frac{b_5}{5}
\left(\begin{array}{ccc} \ell&\ell+2&2\\ 1&-1&0 \end{array}\right)+
\frac{b_6}{5}\sqrt{\frac{3}{2}} 
\left(\begin{array}{ccc} \ell&\ell+2&2\\ 1&1&-2 \end{array}\right)
\right]I^{\ell,\ell+2,2},
\label{K_stelzer}
\end{eqnarray}
\end{widetext}
and $I^{\ell_1,\ell_2,\ell}$ are radial integrals over the expansion coefficients
of the Mayer $f$-function
\begin{equation}
\label{rad_int}
I^{\ell_1,\ell_2,\ell} = \int dr r^4 f^{\ell_1, \ell_2, \ell}(r).
\end{equation}
Expression (\ref{elastic_stelzer}) is the same as 
obtained before by Stelzer {\it et al} \cite{stelzer.j:1995.a}, 
except that we used $3j$ symbols instead of Clebsch-Gordan coefficients 
and a different normalization of the single-particle density.
To simplify (\ref{K_stelzer}) we use the relation between
the $3j$ symbols 
\begin{equation}
\sqrt{\frac{3}{2}}
\left(\begin{array}{ccc} \ell&\ell+2&2\\ 1&1&-2 \end{array}\right) = 
\frac{1}{2}
\left(\begin{array}{ccc} \ell&\ell+2&2\\ 1&-1&0 \end{array}\right),
\end{equation} 
and take into account that sums with $b_3$, $b_5$ and $b_4$, $b_6$
are identical, since 
\begin{equation}
\nonumber
\left(\begin{array}{ccc} \ell_1&\ell_2&2\\ 1&-1&0 \end{array}\right) = 
(-1)^{\ell_1+\ell_2}
\left(\begin{array}{ccc} \ell_2&\ell_1&2\\ -1&1&0 \end{array}\right) =
\left(\begin{array}{ccc} \ell_2&\ell_1&2\\ 1&-1&0 \end{array}\right),
\end{equation} 
and $I^{\ell_1,\ell_2,2} = I^{\ell_2,\ell_1,2}$.
Performing simplifications and taking into account explicit expressions
for the $3j$ coefficients \cite{edmonds.ar:1974}
we obtain expression (\ref{elastic_constants}).

\section{Expansion of the kernel}
\label{sec:hermite}
From the numerical data we found that the kernel of the integral equation 
$A_{221}(z)$ can be accurately approximated with a Gaussian function
\begin{equation}
A_{2,2,1}(z) = - \frac{V_{22}}{4\pi \sqrt{\pi} l} \exp\left[-( z/l )^2\right],
\end{equation}
where we took into account that 
$- 4\pi\int_{-\infty}^\infty A_{2,2,1}(z)dz = 
V_{22}$. Here $l$ is a geometrical parameter which depends only on the 
elongation of the molecules. 

Using the generating function for the Hermite polynomials 
\cite{gradshtein:1994} 
\begin{equation}
\exp(-t^2 + 2tx) = \sum_{n=0}^\infty H_n(x)\frac{t^n}{n!}
\end{equation}
one can write the kernel in a separable form
\begin{eqnarray}
A_{2,2,1}(z_2-z_1) = - \frac{V_{22}}{4\pi \sqrt{\pi} l} 
\exp\left[-( z_2/l )^2\right]  
\nonumber \times \\
\sum_{n=0}^\infty \frac{1}{n!}
\left(\frac{z_1}{l}\right)^n
H_n\left(z_2/l \right),
\end{eqnarray}
similar to the general case we used in sec.~\ref{sec:onsager_anch}.

Using this approximation to the kernel one can show that the expression 
for the anchoring coefficient (\ref{lambda_ideal}) reads
\begin{equation}
\lambda = l / \sqrt{\pi}.
\end{equation}
For ellipsoids with elongation $e = 15$, we found $l \approx 5.46b$. This
results in an anchoring coefficient $\lambda \approx 3.1b$ which
is much lower than the actual value of the anchoring in the system with 
the density and order parameter variations at the surfaces. Hence, 
changes in the density, order parameter, and, as a result, in the 
director profile in the interface region cannot be neglected 
(see also sec.~\ref{sec:onsager_slab}).


\begin{thebibliography}{33}
\expandafter\ifx\csname natexlab\endcsname\relax\def\natexlab#1{#1}\fi
\expandafter\ifx\csname bibnamefont\endcsname\relax
  \def\bibnamefont#1{#1}\fi
\expandafter\ifx\csname bibfnamefont\endcsname\relax
  \def\bibfnamefont#1{#1}\fi
\expandafter\ifx\csname citenamefont\endcsname\relax
  \def\citenamefont#1{#1}\fi
\expandafter\ifx\csname url\endcsname\relax
  \def\url#1{\texttt{#1}}\fi
\expandafter\ifx\csname urlprefix\endcsname\relax\def\urlprefix{URL }\fi
\providecommand{\bibinfo}[2]{#2}
\providecommand{\eprint}[2][]{\url{#2}}

\bibitem[{\citenamefont{de~Gennes and Prost}(1995)}]{degennes.pg:1995.a}
\bibinfo{author}{\bibfnamefont{P.~G.} \bibnamefont{de~Gennes}}
  \bibnamefont{and} \bibinfo{author}{\bibfnamefont{J.}~\bibnamefont{Prost}},
  \emph{\bibinfo{title}{The Physics of Liquid Crystals}}
  (\bibinfo{publisher}{Clarendon Press}, \bibinfo{address}{Oxford},
  \bibinfo{year}{1995}), \bibinfo{edition}{second, paperback} ed.

\bibitem[{\citenamefont{Rapini and Papoular}(1969)}]{rapini.a:1969.a}
\bibinfo{author}{\bibfnamefont{A.}~\bibnamefont{Rapini}} \bibnamefont{and}
  \bibinfo{author}{\bibfnamefont{M.}~\bibnamefont{Papoular}},
  \bibinfo{journal}{J. de Physique Colloque} \textbf{\bibinfo{volume}{30}},
  \bibinfo{pages}{C4} (\bibinfo{year}{1969}).

\bibitem[{\citenamefont{Subacius et~al.}(1995)\citenamefont{Subacius,
  Pergamenshchik, and Lavrentovich}}]{subacius.d:1995.a}
\bibinfo{author}{\bibfnamefont{D.}~\bibnamefont{Subacius}},
  \bibinfo{author}{\bibfnamefont{V.~M.} \bibnamefont{Pergamenshchik}},
  \bibnamefont{and} \bibinfo{author}{\bibfnamefont{O.~D.}
  \bibnamefont{Lavrentovich}}, \bibinfo{journal}{Appl.\ Phys.\ Lett.}
  \textbf{\bibinfo{volume}{67}}, \bibinfo{pages}{214} (\bibinfo{year}{1995}).

\bibitem[{\citenamefont{Yokoyama and van Sprang}(1985)}]{yokoyama.h:1985.a}
\bibinfo{author}{\bibfnamefont{H.}~\bibnamefont{Yokoyama}} \bibnamefont{and}
  \bibinfo{author}{\bibfnamefont{H.~A.} \bibnamefont{van Sprang}},
  \bibinfo{journal}{J. Appl.\ Phys.} \textbf{\bibinfo{volume}{57}},
  \bibinfo{pages}{4520} (\bibinfo{year}{1985}).

\bibitem[{\citenamefont{Nastishin et~al.}(1999)\citenamefont{Nastishin, Polak,
  Shiyanovskii, Bodnar, and Lavrentovich}}]{nastishin.y.a:1999.a}
\bibinfo{author}{\bibfnamefont{Y.~A.} \bibnamefont{Nastishin}},
  \bibinfo{author}{\bibfnamefont{R.~D.} \bibnamefont{Polak}},
  \bibinfo{author}{\bibfnamefont{S.~V.} \bibnamefont{Shiyanovskii}},
  \bibinfo{author}{\bibfnamefont{V.~H.} \bibnamefont{Bodnar}},
  \bibnamefont{and} \bibinfo{author}{\bibfnamefont{O.~D.}
  \bibnamefont{Lavrentovich}}, \bibinfo{journal}{J. Appl.\ Phys.}
  \textbf{\bibinfo{volume}{86}}, \bibinfo{pages}{4199} (\bibinfo{year}{1999}).

\bibitem[{\citenamefont{Mertelj and \v{C}opi\v{c}}(2000)}]{mertelj.a:2000.a}
\bibinfo{author}{\bibfnamefont{A.}~\bibnamefont{Mertelj}} \bibnamefont{and}
  \bibinfo{author}{\bibfnamefont{M.}~\bibnamefont{\v{C}opi\v{c}}},
  \bibinfo{journal}{Phys.\ Rev.\ E} \textbf{\bibinfo{volume}{61}},
  \bibinfo{pages}{1622} (\bibinfo{year}{2000}).

\bibitem[{\citenamefont{Tjipto-Margo and
  Sullivan}(1988)}]{tjiptomargo.b:1988.a}
\bibinfo{author}{\bibfnamefont{B.}~\bibnamefont{Tjipto-Margo}}
  \bibnamefont{and} \bibinfo{author}{\bibfnamefont{D.}~\bibnamefont{Sullivan}},
  \bibinfo{journal}{J. Chem.\ Phys.} \textbf{\bibinfo{volume}{88}},
  \bibinfo{pages}{6620} (\bibinfo{year}{1988}).

\bibitem[{\citenamefont{Teixeira et~al.}(2001)\citenamefont{Teixeira,
  Chrzanowska, Wall, and Cleaver}}]{teixeira.pic:2001.a}
\bibinfo{author}{\bibfnamefont{P.~I.~C.} \bibnamefont{Teixeira}},
  \bibinfo{author}{\bibfnamefont{A.}~\bibnamefont{Chrzanowska}},
  \bibinfo{author}{\bibfnamefont{G.~D.} \bibnamefont{Wall}}, \bibnamefont{and}
  \bibinfo{author}{\bibfnamefont{D.~J.} \bibnamefont{Cleaver}},
  \bibinfo{journal}{Molec.\ Phys.} \textbf{\bibinfo{volume}{99}},
  \bibinfo{pages}{889} (\bibinfo{year}{2001}).

\bibitem[{\citenamefont{Stelzer et~al.}(1997)\citenamefont{Stelzer, Longa, and
  Trebin}}]{stelzer.j:1997.b}
\bibinfo{author}{\bibfnamefont{J.}~\bibnamefont{Stelzer}},
  \bibinfo{author}{\bibfnamefont{L.}~\bibnamefont{Longa}}, \bibnamefont{and}
  \bibinfo{author}{\bibfnamefont{H.~R.} \bibnamefont{Trebin}},
  \bibinfo{journal}{Phys.\ Rev.\ E} \textbf{\bibinfo{volume}{55}},
  \bibinfo{pages}{7085} (\bibinfo{year}{1997}).


\bibitem[{\citenamefont{Phuong et~al.}(2001)\citenamefont{Phuong, Germano, and
  Schmid}}]{phuong.n.h:2001.a}
\bibinfo{author}{\bibfnamefont{N.~H.} \bibnamefont{Phuong}},
  \bibinfo{author}{\bibfnamefont{G.}~\bibnamefont{Germano}}, \bibnamefont{and}
  \bibinfo{author}{\bibfnamefont{F.}~\bibnamefont{Schmid}},
  \bibinfo{journal}{J. Chem.\ Phys.} \textbf{\bibinfo{volume}{115}},  
  \bibinfo{pages}{7227} (\bibinfo{year}{2001}).

\bibitem[{\citenamefont{Ska\v{c}ej et~al.}(1997)\citenamefont{Ska\v{c}ej,
  Pergamenshchik, Alexe-Ionescu, Barbero, and \v{Z}umer}}]{skacej.g:1997.a}
\bibinfo{author}{\bibfnamefont{G.}~\bibnamefont{Ska\v{c}ej}},
  \bibinfo{author}{\bibfnamefont{V.~M.} \bibnamefont{Pergamenshchik}},
  \bibinfo{author}{\bibfnamefont{A.~L.} \bibnamefont{Alexe-Ionescu}},
  \bibinfo{author}{\bibfnamefont{G.}~\bibnamefont{Barbero}}, \bibnamefont{and}
  \bibinfo{author}{\bibfnamefont{S.}~\bibnamefont{\v{Z}umer}},
  \bibinfo{journal}{Phys.\ Rev.\ E} \textbf{\bibinfo{volume}{56}},
  \bibinfo{pages}{571} (\bibinfo{year}{1997}).

\bibitem[{\citenamefont{Priezjev and Pelcovits}(2000)}]{priezjev.n:2000.a}
\bibinfo{author}{\bibfnamefont{N.}~\bibnamefont{Priezjev}} \bibnamefont{and}
  \bibinfo{author}{\bibfnamefont{R.~A.} \bibnamefont{Pelcovits}},
  \bibinfo{journal}{Phys.\ Rev.\ E} \textbf{\bibinfo{volume}{62}},
  \bibinfo{pages}{6734} (\bibinfo{year}{2000}).

\bibitem[{\citenamefont{Onsager}(1949)}]{onsager.l:1949.a}
\bibinfo{author}{\bibfnamefont{L.}~\bibnamefont{Onsager}},
  \bibinfo{journal}{Ann.\ N. Y. Acad.\ Sci.} \textbf{\bibinfo{volume}{51}},
  \bibinfo{pages}{627} (\bibinfo{year}{1949}).

\bibitem[{\citenamefont{Evans}(1992)}]{evans.r:1992.a}
\bibinfo{author}{\bibfnamefont{R.}~\bibnamefont{Evans}}, in
  \emph{\bibinfo{booktitle}{Fundamentals of Inhomogeneous Fluids}}, edited by
  \bibinfo{editor}{\bibfnamefont{D.}~\bibnamefont{Henderson}}
  (\bibinfo{publisher}{Dekker}, \bibinfo{address}{New York},
  \bibinfo{year}{1992}), chap.~\bibinfo{chapter}{3}, pp.
  \bibinfo{pages}{85--175}.

\bibitem[{\citenamefont{Gray and Gubbins}(1984)}]{gray.cg:1984.a}
\bibinfo{author}{\bibfnamefont{C.~G.} \bibnamefont{Gray}} \bibnamefont{and}
  \bibinfo{author}{\bibfnamefont{K.~E.} \bibnamefont{Gubbins}},
  \emph{\bibinfo{title}{Theory of molecular fluids. 1. {F}undamentals}}
  (\bibinfo{publisher}{Clarendon Press}, \bibinfo{address}{Oxford},
  \bibinfo{year}{1984}).

\bibitem[{\citenamefont{Press et~al.}(1992)\citenamefont{Press, Flannery,
  Teukolsky, and Vetterling}}]{press.wh:1992.a}
\bibinfo{author}{\bibfnamefont{W.~H.} \bibnamefont{Press}},
  \bibinfo{author}{\bibfnamefont{B.~P.} \bibnamefont{Flannery}},
  \bibinfo{author}{\bibfnamefont{S.~A.} \bibnamefont{Teukolsky}},
  \bibnamefont{and} \bibinfo{author}{\bibfnamefont{W.~T.}
  \bibnamefont{Vetterling}}, \emph{\bibinfo{title}{Numerical Recipes in
  Fortran}} (\bibinfo{publisher}{Cambridge University Press},
  \bibinfo{address}{Cambridge}, \bibinfo{year}{1992}), \bibinfo{edition}{2nd}
  ed.

\bibitem[{\citenamefont{Poniewierski and Stecki}(1979)}]{poniewierski.a:1979.a}
\bibinfo{author}{\bibfnamefont{A.}~\bibnamefont{Poniewierski}}
  \bibnamefont{and} \bibinfo{author}{\bibfnamefont{J.}~\bibnamefont{Stecki}},
  \bibinfo{journal}{Molec.\ Phys.} \textbf{\bibinfo{volume}{38}},
  \bibinfo{pages}{1931} (\bibinfo{year}{1979}).

\bibitem[{\citenamefont{Lipkin et~al.}(1985)\citenamefont{Lipkin, Rice, and
  Mohanty}}]{lipkin.md:1985.a}
\bibinfo{author}{\bibfnamefont{M.~D.} \bibnamefont{Lipkin}},
  \bibinfo{author}{\bibfnamefont{S.~A.} \bibnamefont{Rice}}, \bibnamefont{and}
  \bibinfo{author}{\bibfnamefont{U.}~\bibnamefont{Mohanty}},
  \bibinfo{journal}{J. Chem.\ Phys.} \textbf{\bibinfo{volume}{82}},
  \bibinfo{pages}{472} (\bibinfo{year}{1985}).

\bibitem[{\citenamefont{Yokoyama}(1997)}]{yokoyama.h:1997.a}
\bibinfo{author}{\bibfnamefont{H.}~\bibnamefont{Yokoyama}},
  \bibinfo{journal}{Phys.\ Rev.\ E} \textbf{\bibinfo{volume}{55}},
  \bibinfo{pages}{2938} (\bibinfo{year}{1997}).

\bibitem[{\citenamefont{Stelzer et~al.}(1995)\citenamefont{Stelzer, Longa, and
  Trebin}}]{stelzer.j:1995.a}
\bibinfo{author}{\bibfnamefont{J.}~\bibnamefont{Stelzer}},
  \bibinfo{author}{\bibfnamefont{L.}~\bibnamefont{Longa}}, \bibnamefont{and}
  \bibinfo{author}{\bibfnamefont{H.~R.} \bibnamefont{Trebin}},
  \bibinfo{journal}{J. Chem.\ Phys.} \textbf{\bibinfo{volume}{103}},
  \bibinfo{pages}{3098} (\bibinfo{year}{1995}).

\bibitem[{\citenamefont{Somoza and Tarazona}(1991)}]{somoza.am:1991.a}
\bibinfo{author}{\bibfnamefont{A.~M.} \bibnamefont{Somoza}} \bibnamefont{and}
  \bibinfo{author}{\bibfnamefont{P.}~\bibnamefont{Tarazona}},
  \bibinfo{journal}{Molec.\ Phys.} \textbf{\bibinfo{volume}{72}},
  \bibinfo{pages}{911} (\bibinfo{year}{1991}).

\bibitem[{\citenamefont{Andrienko et~al.}(2000)\citenamefont{Andrienko,
  Germano, and Allen}}]{andrienko.d:2000.d}
\bibinfo{author}{\bibfnamefont{D.}~\bibnamefont{Andrienko}},
  \bibinfo{author}{\bibfnamefont{G.}~\bibnamefont{Germano}}, \bibnamefont{and}
  \bibinfo{author}{\bibfnamefont{M.~P.} \bibnamefont{Allen}},
  \bibinfo{journal}{Phys.\ Rev.\ E} \textbf{\bibinfo{volume}{62}},
  \bibinfo{pages}{6688} (\bibinfo{year}{2000}).

\bibitem[{\citenamefont{Allen and Frenkel}(1988)}]{allen.mp:1988.a}
\bibinfo{author}{\bibfnamefont{M.~P.} \bibnamefont{Allen}} \bibnamefont{and}
  \bibinfo{author}{\bibfnamefont{D.}~\bibnamefont{Frenkel}},
  \bibinfo{journal}{Phys.\ Rev.\ A} \textbf{\bibinfo{volume}{37}},
  \bibinfo{pages}{1813} (\bibinfo{year}{1988}).

\bibitem[{\citenamefont{Allen and Frenkel}(1990)}]{allen.mp:1990.a}
\bibinfo{author}{\bibfnamefont{M.~P.} \bibnamefont{Allen}} \bibnamefont{and}
  \bibinfo{author}{\bibfnamefont{D.}~\bibnamefont{Frenkel}},
  \bibinfo{journal}{Phys.\ Rev.\ A} \textbf{\bibinfo{volume}{42}},
  \bibinfo{pages}{3641} (\bibinfo{year}{1990}), \bibinfo{note}{erratum}.

\bibitem[{\citenamefont{Allen}(1999)}]{allen.mp:1999.a}
\bibinfo{author}{\bibfnamefont{M.~P.} \bibnamefont{Allen}},
  \bibinfo{journal}{Molec.\ Phys.} \textbf{\bibinfo{volume}{96}},
  \bibinfo{pages}{1391} (\bibinfo{year}{1999}).

\bibitem[{\citenamefont{Forster}(1975)}]{forster.d:1975.a}
\bibinfo{author}{\bibfnamefont{D.}~\bibnamefont{Forster}},
  \emph{\bibinfo{title}{Hydrodynamic Fluctuations, Broken Symmetry and
  Correlation Functions}}, vol.~\bibinfo{volume}{47} of
  \emph{\bibinfo{series}{Frontiers in Physics}} (\bibinfo{publisher}{Benjamin},
  \bibinfo{address}{Reading}, \bibinfo{year}{1975}).

\bibitem[{\citenamefont{Frenkel et~al.}(1984)\citenamefont{Frenkel, Mulder, and
  McTague}}]{frenkel.d:1984.a}
\bibinfo{author}{\bibfnamefont{D.}~\bibnamefont{Frenkel}},
  \bibinfo{author}{\bibfnamefont{B.~M.} \bibnamefont{Mulder}},
  \bibnamefont{and} \bibinfo{author}{\bibfnamefont{J.~P.}
  \bibnamefont{McTague}}, \bibinfo{journal}{Phys.\ Rev.\ Lett.}
  \textbf{\bibinfo{volume}{52}}, \bibinfo{pages}{287} (\bibinfo{year}{1984}).

\bibitem[{\citenamefont{Frenkel and Mulder}(1985)}]{frenkel.d:1985.a}
\bibinfo{author}{\bibfnamefont{D.}~\bibnamefont{Frenkel}} \bibnamefont{and}
  \bibinfo{author}{\bibfnamefont{B.~M.} \bibnamefont{Mulder}},
  \bibinfo{journal}{Molec.\ Phys.} \textbf{\bibinfo{volume}{55}},
  \bibinfo{pages}{1171} (\bibinfo{year}{1985}).

\bibitem[{\citenamefont{Tjipto-Margo et~al.}(1992)\citenamefont{Tjipto-Margo,
  Evans, Allen, and Frenkel}}]{tjiptomargo.b:1992.a}
\bibinfo{author}{\bibfnamefont{B.}~\bibnamefont{Tjipto-Margo}},
  \bibinfo{author}{\bibfnamefont{G.~T.} \bibnamefont{Evans}},
  \bibinfo{author}{\bibfnamefont{M.~P.} \bibnamefont{Allen}}, \bibnamefont{and}
  \bibinfo{author}{\bibfnamefont{D.}~\bibnamefont{Frenkel}},
  \bibinfo{journal}{J. Phys.\ Chem.} \textbf{\bibinfo{volume}{96}},
  \bibinfo{pages}{3942} (\bibinfo{year}{1992}).

\bibitem[{\citenamefont{Allen}(1993)}]{allen.mp:1993.g}
\bibinfo{author}{\bibfnamefont{M.~P.} \bibnamefont{Allen}},
  \bibinfo{journal}{Phil.\ Trans.\ Roy.\ Soc.\ A}
  \textbf{\bibinfo{volume}{344}}, \bibinfo{pages}{323} (\bibinfo{year}{1993}),
  \bibinfo{note}{theme Issue: Understanding Self-assembly and Organization in
  Liquid Crystals}.

\bibitem[{\citenamefont{Allen et~al.}(1993)\citenamefont{Allen, Evans, Frenkel,
  and Mulder}}]{allen.mp:1993.h}
\bibinfo{author}{\bibfnamefont{M.~P.} \bibnamefont{Allen}},
  \bibinfo{author}{\bibfnamefont{G.~T.} \bibnamefont{Evans}},
  \bibinfo{author}{\bibfnamefont{D.}~\bibnamefont{Frenkel}}, \bibnamefont{and}
  \bibinfo{author}{\bibfnamefont{B.}~\bibnamefont{Mulder}},
  \bibinfo{journal}{Adv.\ Chem.\ Phys.} \textbf{\bibinfo{volume}{86}},
  \bibinfo{pages}{1} (\bibinfo{year}{1993}).

\bibitem[{\citenamefont{Edmonds}(1974)}]{edmonds.ar:1974}
\bibinfo{author}{\bibfnamefont{A.~R.} \bibnamefont{Edmonds}},
  \emph{\bibinfo{title}{Angular momentum in quantum mechanics}}
  (\bibinfo{publisher}{Princeton University Press},
  \bibinfo{address}{Princeton, New Jersey}, \bibinfo{year}{1974}).

\bibitem[{\citenamefont{Gradshtein and Ryzhik}(1994)}]{gradshtein:1994}
\bibinfo{author}{\bibfnamefont{I.~S.} \bibnamefont{Gradshtein}}
  \bibnamefont{and} \bibinfo{author}{\bibfnamefont{I.~M.}
  \bibnamefont{Ryzhik}}, \emph{\bibinfo{title}{Tables of integrals, series and
  products}} (\bibinfo{publisher}{Academic Press}, \bibinfo{address}{Boston},
  \bibinfo{year}{1994}).

\end{thebibliography}

%
\end{document}